\documentclass[%
 reprint,
 amsmath,amssymb,
 aps,
pra,
]{revtex4-2}

\usepackage{graphicx}
\usepackage{dcolumn}
\usepackage{bm}

\usepackage{multirow}
\usepackage{color}
\usepackage{tikz}
\definecolor{green}{RGB}{0,128,0}
\definecolor{yellow}{RGB}{255,190,0}
\begin{document}

\preprint{APS/123-QED}
\title{Helium Bubbles in Liquid Lead--Lithium Solutions: Pressure Inhomogeneities at Interfaces and Non-Ideal Mixture Effects}
\thanks{Ministerios, CSN/ARGOS, UPC and EUROfusion...}%

\author{Edgar Alvarez-Galera}
 \altaffiliation[Campus Diagonal Nord B4-201, ]{08034 Barcelona, Catalonia, Spain}
 \email{edgar.alvarez.galera@upc.edu}
\author{Jordi Mart{\'i}}%
\altaffiliation[Campus Diagonal Nord B5-209, ]{08034 Barcelona, Catalonia, Spain}
 \email{jordi.marti@upc.edu}
\author{Lluis Batet}
\altaffiliation[Campus Diagonal Sud H-32/02, ]{Diagonal 647, 08028 Barcelona, Catalonia, Spain}
\affiliation{%
Department of Physics, Polytechnic University of Catalonia--Barcelona Tech.\\
C. Jordi Girona, 1-3 08034 Barcelona \textbackslash\textbackslash
}%

\date{\today}

\begin{abstract}
    The extremely low solubility of helium in liquid metals may lead to rapid supersaturation, promoting spontaneous {formation} of helium bubbles by nucleation. Once nucleated, the stability of these bubbles is governed by the properties of the helium--liquid metal interface. In particular, interfacial tension between the immiscible phases controls bubble interactions and induces local pressure inhomogeneities.
    This work is motivated by the need of a better understanding of helium bubble formation in liquid Pb--Li alloys, which are of particular relevance for {the design of breeding blankets in the future nuclear fusion reactors}. We employ classical molecular dynamics simulations to investigate helium segregation in a range of lead--lithium systems, including the limiting cases of pure lead and pure lithium.
    Changes in local pressure are evaluated from direct mechanical calculations, enabling the characterization of interfacial properties. Interfacial tension and radius of the bubble are subsequently determined across multiple thermodynamic conditions, spanning temperatures starting near the melting points of the constituent metals up to 1021 K. The impact of curvature and composition of the alloy on the interfacial behaviour are also investigated.
\begin{description}
\item[Usage]
    Secondary publications and information retrieval purposes.
\item[Structure]
    The paper is organized as follows: In Sec.~\ref{sec:MODELS} we present the force-field models employed in this work.
    In Sec.~\ref{sec:THEORY_and_METHODS} we review the MD theory of pressure tensors and interfacial tension calculations. 
    The main results and discussion are reported in Sec.~\ref{sec:RESULTS} and some final remarks and conclusions are given in Sec.~\ref{sec:CONCLUSIONS}.
\end{description}
\end{abstract}

\maketitle

\section{\label{sec:Introduction}Introduction}
    The formation of helium nanobubbles and {the} solubility {of helium} in liquid metals (LM) and in liquid metal alloys (LMA) has been a subject of experimental and theoretical interest over the past decades~\cite{borgstedt2001iupac,
    conrad1991irradiation,
    conrad1994libretto,kordavc2017helium,
    fraile2020volume,
    haas2021review,
    alvarez2023nucleation,
    kekrt2023concept,
    saraswat2025comprehensive,
    gottfriedova6232758experimental}.
    Among several LMA technological applications, the proposal of the lead--lithium eutectic (LLE) alloy as breeding material~\cite{kordavc2017helium} in the framework of nuclear fusion projects ITER~\cite{iter-web} and DEMO~\cite{demo-web}
    is of outmost importance.

    Helium, released as a by-product during the tritium breeding, segregates due to its predominant repulsive interactions with the surrounding media.
    Understanding the mechanisms behind gas segregation in liquid solutions is crucial for optimizing tritium extraction in fusion breeding blankets and holds fundamental importance.
    Despite the phenomenon of nucleation has been addressed from an elemental perspective during the last century~\cite{volmer1926keimbildung, becker1935kinetische, gunton1999homogeneous}, a complete and unified theory is still lacking, and it continues attracting scientific interest~\cite{llamas2026free}.
    Nucleation plays a key role in diverse processes, ranging from droplet condensation~\cite{valencia2006kinetics}, ice nucleation~\cite{montero2026non}, aerosol sciences~\cite{vadgama2025thermodynamical, ICNAA2025_Abstracts},  and cavitation~\cite{bazhirov2008cavitation}, to the aforementioned formation of helium nanobubbles~\cite{batet2011numeric, fraile2020volume, marti2022nucleation, alvarez2023nucleation, al2024stochastic}.
    In this work, helium bubble formation is interpreted as a cavitation process driven by the supersaturation of the noble gas within the liquid metal solution, triggered by segregation
    due to repulsive interactions between helium and the surrounding LM, rather than aggregation due to attractive forces among the solute atoms.

    This work employs classical molecular dynamics (MD) to investigate the behaviour of interfacial forces in helium/lead-lithium liquid alloys (He/LLLA) interfaces from an atomistic perspective. The study covers both{: (i)} planar; and, (ii) curved interfaces.
    Analysing all-atom trajectories in MD simulations enables the calculation of normal and tangential stress tensor components. These components, given the specific geometry, are:
    (i) profiles varying along the Cartesian coordinate perpendicular to the interface (i.e., $z$-dependent)~\cite{kirkwood1949statistical}; and, (ii) radial profiles (i.e., $r$-dependent)~\cite{thompson1984molecular}.

    We pay special attention to changes in the interfacial tension induced by modifying the atomic composition of the LM solvent. Many properties related to lead--lithium alloys are known to show thermodynamic non-ideal mixture characteristics~\cite{buxbaum1984chemical}.
    In a previous study~\cite{alvarez2025henry} we demonstrated that the same force field employed in this work~\cite{al2023parametrization, sladek2014ab, sheng2021development} reproduces a notable peculiarity for both solvent packing factors and helium solubilities across a wide range of lead to lithium ratios along the 1000~K isotherm, coherent with an experimentally-determined divergent behaviour of bulk properties such as volumetric thermal expansion coefficients~\cite{ruppersberg1976density, khairulin2017volumetric} or heat capacities~\cite{saar1987calculation, fraile2014interatomic}. 
    In the present work we explore how pressure tensors and interfacial properties of He/LLLA, including how the radius of the bubbles
    are affected by the composition of the LMA. These properties are studied {at nominal conditions of $T\approx1021.4$~K and $P\approx1$~bar, {showing} reasonable ensemble fluctuations} when varying the ratio between lead and lithium atoms for a few selected states.

    The paper is organized as follows: In Sec.~\ref{sec:MODELS} we present the force-field models employed in this work.
    In Sec.~\ref{sec:THEORY_and_METHODS} we review the MD theory of pressure tensors and interfacial tension calculations. 
    The main results and discussion are reported in Sec.~\ref{sec:RESULTS} and some final remarks and conclusions are given in Sec.~\ref{sec:CONCLUSIONS}.

\section{\label{sec:MODELS}Models}
    \subsection{{Force fields}}
        Atomistic trajectories are {governed} by a classical Hamiltonian,
        {typically decomposed into} kinetic and potential terms. The latter includes: 
        \paragraph{EAM interactions for LLLA~\cite{al2023parametrization, belashchenko2019inclusion}}
            {The Embedded Atom Model} (EAM) description was used in Refs.~\citenum{al2023parametrization,belashchenko2019inclusion,fraile2014interatomic} to model Li--Pb molten alloys, {capturing} the effective contribution of free-electron clouds~\cite{daw1993embedded,daeneke2018liquid} and the large non-idealities in solution thermodynamics~\cite{buxbaum1984chemical, alvarez2025henry}. The existence of a peculiarity in the 1000~K isotherm around the Pb-Li4 (Li80\%-Pb20\%) composition was discussed in Ref.~\citenum{alvarez2025henry}, 
            {where it} was shown to be in accordance with experimental observations~\cite{okamoto1993li, saar1987calculation, khairulin2017volumetric}.
            \paragraph{Helium pairwise interactions~\cite{sheng2021development, sladek2014ab}}
            {The interactions of helium} with other atoms can be considered to be purely pair-decomposable. As a noble gas, helium atoms have a closed and stable shell. 
            This suggests that free electrons do not affect interactions in condensed phases. Consequently, the Pb--He~\cite{sladek2014ab} and the Li--He, He--He interactions~\cite{sheng2021development} can be directly transferred from vacuum conditions {where} those potentials were built.
            In summary, the total Hamiltonian of the system reads as:
        \begin{eqnarray}
            \mathcal{H} &=& 
                \sum_{\alpha \in \mathbb{T}} \frac{1}{2 m_\alpha} \sum_{i=1}^{N_\alpha} \mathbf{p}_i^\alpha \cdot \mathbf{p}_i^\alpha
            \nonumber\\
            &&
                + \sum_{\alpha \in \mathbb{B}} \sum_{i=1}^{N_\alpha} \Bigg[\Phi_\mathrm{ \alpha} (\varrho_i^\alpha) 
                +\frac{1}{2} \sum_{\beta \in \mathbb{B}} \sum_{j=1}^{N_\beta}  \varphi_{\alpha-\beta}(r_{ij}^{\alpha \beta})\Bigg]
                \nonumber \\
                && + \frac{1}{2}\sum_{i=1}^{N_\mathrm{He}} \sum_{\gamma \in \mathbb{T}} \sum_{j=1}^{N_\gamma} \varphi_{\mathrm{ He} - \gamma} (r_{ij}^{\mathrm{ He} - \gamma})
                \ ,
            \label{eq:Hamiltonian}
        \end{eqnarray}
        {where $\mathbb{B} \equiv \left\{ \rm Li, Pb \right\}$ and $\mathbb{T} \equiv \left\{ \rm Li, Pb, He \right\}$ denote the binary (LM, excluding He) and ternary mixtures, respectively}; $m_\alpha$ and $\mathbf{p}_i^\alpha$ {are} the mass and momentum of an 
        atom $i$ of species $\alpha$; $\varrho_i^\alpha \equiv \sum_{\beta\in\mathbb{B}} (1-\delta_{ij}\delta^{\alpha\beta}) \ n_{ij}^{\alpha\beta}$ {is} the effective electron-cloud density {at} the position of a LM atom due to the presence of other neighbouring LM atoms, expressed as the superposition of pairwise electron densities, ${n_{ij}^{\alpha\beta} \equiv} n(r_{ij}^{\alpha\beta})$; and, {as usual,} $\varphi_{ij}^{\alpha\beta} \equiv \varphi_{\alpha-\beta} (r_{ij}^{\alpha\beta})$ and $r_{ij}^{\alpha\beta} \equiv |\mathbf{r}_j^\beta - \mathbf{r}_i^\alpha|$ {are} the pair potential and distance between two atoms $i$ and $j$, of species $\alpha$ and $\beta$, respectively. The {reliability} of this force field was properly discussed in terms of its accuracy and transferability in Ref.~\citenum{alvarez2025henry}.

    \subsection{{Technical details}}
        Positions and velocities are discretized in time, with a time step of $\Delta 
        t = 2$~fs, and the integration {algorithm} is the standard velocity-Verlet~\cite{swope1982computer}. 
        Isothermal--isobaric (NPT) trajectories are generated using the standard Nos{\'e}--Hoover barostat~\cite{nose1984unified,hoover1985canonical,martyna1994constant}. 
        All simulations include periodic boundary conditions (PBC) along the three directions of space.
        The sizes of simulation boxes are chosen as a trade-off between computational effort and reduction of finite size effects. 

\section{\label{sec:THEORY_and_METHODS}Theory and Methods}
    \subsection{Thermodynamics of bubbles}
        As a starting point, formation of helium clusters in a liquid metal solution can be interpreted as a nucleation process. Helium and supersaturated lead--lithium alloys are regarded as two immiscible phases, 
        that remain separated after the apparition of an interface.
        A hypothetical supersaturated solution containing $N_{\rm He}$ helium atoms {initially dissolved into the LM bath} is metastable compared to a configuration in which helium atoms segregate and form clusters, due to the extremely low solubility of helium~\cite{borgstedt2001iupac, alvarez2024henry, alvarez2025henry}. The thermodynamic driving force for this transition arises from the supersaturation of helium in the host liquid metal, which produces an effective bulk free-energy gain associated with phase separation.
        
        Within the framework of classical nucleation theory, the formation of a cluster involves two competing contributions to the free energy. First, the segregation of helium atoms into a dense phase lowers the bulk free energy of the system. Second,
        creating an interface between the helium-rich cluster and the surrounding liquid metal imposes an energetic penalty proportional to the interfacial area due to interfacial tension. As a result, the total free energy of cluster formation reflects the competition between a favourable bulk contribution and an unfavourable interfacial contribution.
        
        For a spherical cluster of radius $\mathcal{R}$, the free-energy change associated with its formation can be written as:
        \begin{equation}
        \Delta G(\mathcal{R}) = -\underbrace{\frac{4}{3}\pi \mathcal{R}^3 \Delta g}_{(\Delta G)_{bulk}} + \underbrace{4\pi \mathcal{R}^2 \gamma (\mathcal{R})}_{(\Delta G)_{interf}} ,
        \end{equation}
        where $\gamma$ is the interfacial tension between the helium cluster and the surrounding liquid metal, and $\Delta g \equiv (\beta \ v_0)^{-1} \ln{\Psi}$ is the bulk free-energy difference per unit volume between the two phases, being $\beta$ the thermodynamic beta, $v_0$ the volume of a helium monomer (with $v_0^{-1}$ the density of the bubble), and $\Psi$ denotes the supersaturation ratio, defined as the fraction between the actual pressure of the system and the equilibrium partial pressure of dissolved helium. Assuming that helium is infinitely diluted, Henry's law determines that $\Psi$ can equivalently be expressed as fraction between the actual concentration of dissolved helium, and the equilibrium concentration at the local pressure. 
        As a matter of fact, (i) the bulk term divided by the number of clustered atoms measures the change in chemical potentials between the initial ($\mu_d$) and final {($\mu_c$)} phases, $(\Delta G)_{bulk}/N_{\rm He} = \Delta \mu = \mu_d-\mu_c = \beta^{-1}\ln{\Psi}$ --which determines the thermodynamic driving force due to solute excess-- and (ii) the interfacial term accounts for the work required to form a dividing surface between the nucleated helium and the liquid solution.
        
        Far away from interfaces, helium bubbles and {bulk helium (dissolved, within the LM solution)} are assumed to have uniform densities, {namely} $\rho_{\rm He}^c$ and $\rho_{\rm He}^d$, respectively.
        In contrast, there are excess densities of helium at interfaces
        --assuming extra number of atoms due to adsorption, with interfacial concentration $\Gamma$-- 
        followed by monotonic decreases until reaching bulk helium densities. Bubbles are confined
        within the spherical regions of radius $\mathcal{R}$, with volume $V_{int} = \frac{4}{3} \pi \mathcal{R}^3$ and area $A_{int} = 4 \pi \mathcal{R}^2$, while helium atoms outside these spheres are considered to be within the liquid solution.
        Therefore, the total number of helium atoms within the simulation box can be written as the sum of the partial number of atoms (i) within the bubble, (ii) dissolved in the solution, and (iii) {an} excess quantity at the interface:
        \begin{eqnarray}
            N_{\rm He}^* &=& \int \mathrm{d}^3\mathbf{r} \ \rho_{\rm He}(|\mathbf{r} - \mathbf{r}_{CoM}|) 
            \\
            &=& \rho_{\rm He}^c V_{int} + \rho_{\rm He}^{d} (V-V_{int}) + \Gamma A_{int}  \ ,
        \end{eqnarray}
        where $\rho_{\rm He}(r)$ denotes the helium number density profile with origin at the centre of mass (CoM) of the bubble{, $r \equiv |\mathbf{r} - \mathbf{r}_{CoM}|$}.

        Al-Awad~\cite{awad2025multiscale} discussed the stability of helium nanobubbles (at {$T=600$~K}) surrounded by both pure LLE and the corresponding solution supersaturated in helium, concluding that an `artificial' shielding mechanism (related to PBC) was required in MD simulations to avoid helium partial dissolution.
        However, the location of extra atoms near the interface is not an artifact, rather it possesses a clear physical insight: a few helium atoms are adsorbed to      the interface in order to minimize the total free energy.
        Adsorption refers to the adhesion of helium atoms from the solution to the surface of the cluster, where those extra atoms in the surface do not penetrate inside the bulk of the cluster.
        
        Regarding total density profiles centred in the {CoM} of helium clusters,
        \[
            \rho (r) = \rho_\text{Li} (r) + \rho_\text{Pb} (r) + \rho_\text{He} (r)\ ,
        \]
        we should expect to observe: (i) two plateaus, with densities $\rho_\text{He}^c$ and $\rho_\text{LM} + \rho_\text{He}^d$, corresponding to short ($r\rightarrow0$) and long ($r\rightarrow\infty$) distance limits, respectively; (ii) a depletion of atoms in the interfacial region.

    \subsection{Pressure inhomogeneities and interfacial tension}
        \subsubsection{Pressure tensors}
        \paragraph{Mechanical equilibrium}
        In equilibrium, the pressure tensor {$\mathbf{\Pi}$} and the hydrostatic equilibrium condition under a given orthonormal coordinate system $\{\mathbf{e}_1, \mathbf{e}_2, \mathbf{e}_3\}$ respectively follow as
        \begin{eqnarray}
            \mathbf{\Pi} = \sum_{\mu=1}^3 \sum_{\nu=1}^3 \Pi_{\mu\nu} \mathbf{e}_\mu \otimes \mathbf{e}_\nu \ , \text{ and}
            \\
            \nabla \cdot \mathbf{\Pi} = 0 \ ,
        \end{eqnarray}
        where $\mathbf{e}_\mu$ is the unit vector along the direction $\mu$ and 
        the vector operator nabla is defined as
        $\displaystyle \nabla \equiv \sum_{\mu=1}^3 \mathbf{e}_\mu \frac{\partial}{\partial x_\mu}$.
        The proper choice of basis coordinates depend on the geometry of the interface under study: Cartesian coordinates are suitable for flat interfaces~\cite{kirkwood1949statistical}, while spherical coordinates are appropriate for spherical interfaces (e.g., droplets, bubbles, or cavities)~\cite{thompson1984molecular}.

        \paragraph{Calculation of pressure tensors}
          From the fact that the mechanical approach {to  calculate interfacial tensions involves} pressure tensors, we can derive some useful expressions. 
          For an orthorhombic (rectangular prism) simulation cell of volume $V = L_x L_y L_z$, the virial expression for the pressure tensor is decomposed into kinetic and configurational components~\cite{kirkwood1949statistical, irving1950statistical, buff1951spherical, buff1955spherical, walton1983pressure, tsai1979virial, ikeshoji2003molecular, gloor2005test, gonzalez2005molecular, smith2017towards}:
        \begin{eqnarray}
            \mathbf{P} = \mathbf{P}_{\rm K} + \mathbf{P}_{\mathrm{V}} \ ,
        \end{eqnarray}
        with its components reading as:
        \begin{eqnarray}
            &&\mathbf{e}_\mu \cdot \mathbf{P} \cdot\mathbf{e}_\nu = \langle \Pi_{\mu\nu}\rangle 
            \nonumber\\
            &=& 
            \underbrace{\left\langle \frac{1}{V} \sum_{\alpha, i} m_\alpha (\mathbf{v}_i^\alpha)_\mu (\mathbf{v}_i^\alpha)_\nu  \right\rangle}_\text{kinetic}
            \nonumber\\
            &&+
            \underbrace{\left\langle \frac{1}{2V} \sum_{\alpha, i} \sum_{\beta, j} (\mathbf{r}_{ij}^{\alpha\beta})_{\mu} (\mathbf{f}_{ij}^{\alpha\beta})_{\nu} \right\rangle}_\text{virial or configurational}
            \ .
        \end{eqnarray}
        The diagonal components can be computed from all-atom instantaneous positions and forces, while off-diagonal components (after the ensemble averaging, denoted by $\langle \dots \rangle$) turn 
        out to be zero.
        Vectors $\mathbf{r}_i^\alpha, \mathbf{v}_i^\alpha$ denote the position and velocity of the $i$-th atom (respectively), while $\displaystyle\mathbf{f}_{ij}^{\alpha\beta} = (1-\delta_{ij}\delta^{\alpha\beta}) f_{ij}^{\alpha\beta} \frac{\mathbf{r}_{ij}^{\alpha\beta}}{|\mathbf{r}_{ij}^{\alpha\beta}|}$ is the force {exerted by} $i$ on $j$, and $\mathbf{r}_{ij}^{\alpha\beta} \equiv \mathbf{r}_j^\beta - \mathbf{r}_i^\alpha$ is the displacement vector between the former atoms.
        For the force field {given in (\ref{eq:Hamiltonian})},
        scalar components of pair forces read 
        \[
            f_{ij}^{\alpha\beta} = \begin{cases}
                \displaystyle - \left(\frac{\mathrm{d}\Phi_j^\beta}{\mathrm{d}\varrho_j^\beta}
                \frac{\mathrm{d}n_{ij}^{\alpha\beta}}{\mathrm{d}r_{ij}^{\alpha\beta}}
                + \frac{\mathrm{d}\Phi_i^\alpha}{\mathrm{d}\varrho_i^\alpha} \frac{\mathrm{d}n_{ji}^{\beta\alpha}}{\mathrm{d}r_{ji}^{\beta\alpha}}\right)
                - \frac{\mathrm{d}\varphi_{ij}^{\alpha\beta}}{\mathrm{d}r_{ij}^{\alpha\beta}} \ , \\
                \hfill \text{if } \alpha,\beta \in \{{\rm Li,Pb}\}  \ ; \\
                \displaystyle - \frac{\mathrm{d}\varphi_{ij}^{\alpha\beta}}{\mathrm{d}r_{ij}^{\alpha\beta}} \ , \hfill \text{otherwise}. 
            \end{cases}
        \]
        
        Calculating pressure-related quantities from atomic trajectories is generally a delicate task due to the large fluctuations involved in the process.
        In particular, approaching the interfacial tension from simulations at the nanoscale implies the calculation of fluctuating magnitudes in the position space, not well-defined at a global level~\cite{park2001molecular,caro2013structure,caro2015capillarity}.
        However, after sufficient statistical averaging, the final pressure tensors converge to smooth functions of positions.

        \paragraph{Local pressure tensors}
        In bulk systems, the pressure is spatially homogeneous. However, at the vicinity of interfaces, pressure variations emerge as a result of local interactions between the constituent species.
        The mechanical route to compute the interfacial tension is based on the evaluation of pressure tensor local anisotropies, namely the difference between its normal and tangential components across the interface.
        Instantaneous local values of pressure tensors are evaluated in finite layers of volume $\Delta V = A \ \Delta \xi_N$, where $\Delta \xi_N$ denotes the width of a layer in the direction normal to the interface and $A$ the area of the plane parallel to the interface.
        This discretization allows for the calculation of momentum flux across the defined boundaries of the control volume. 
        The concept of localization is implemented by taking the differential limits $\Delta \xi_N \rightarrow 0$, effectively collapsing the finite slab into a mathematical plane. This transition from a discrete volume to a continuous field allows the pressure to be defined at a specific point $\mathbf{r}$ in space.
        Following the work of Kirkwood and Buff~\cite{kirkwood1949statistical, buff1951spherical, buff1955spherical}, Thompson et al.~\cite{thompson1984molecular}, and Wen et al.~\cite{wen2021molecular}, we express the local components of the pressure tensor along the normal coordinate $\xi_N$ (defined by the symmetry of the interface), as 
        \begin{eqnarray} \label{eq:stress_sym}
            \Pi_{\mu\nu} (\xi_N) &=& \lim_{\Delta V \rightarrow0}
            \frac{\Delta S_{\mu\nu}(\xi_N)}{\Delta V}
            \nonumber\\
            &=& \lim_{\Delta V \rightarrow0} \frac{1}{\Delta V}\Bigg[ \Delta n(\xi_N) \beta^{-1} 
            \nonumber\\
            && + \frac{1}{2} \sum_{\alpha, i} \sum_{\beta, j} [\mathbf{r}_{ij}^{\alpha\beta}]_\mu [\mathbf{f}_{ij}^{\alpha\beta}]_\nu \ {\Delta \Omega}_{ij}(\xi_N) \Bigg]
            \nonumber \\
            &\simeq& 
            \Bigg[\rho(\xi_N) \beta^{-1} 
            \nonumber\\
            &&+ \frac{1}{2} \sum_{\alpha, i} \sum_{\beta, j} [\mathbf{r}_{ij}^{\alpha\beta}]_\mu [\mathbf{f}_{ij}^{\alpha\beta}]_\nu \  \frac{\partial \Omega_{ij}^{\alpha\beta}}{\partial \xi_N} \Bigg]
            \nonumber
            \ ,
        \end{eqnarray}
        where $S_{\mu \nu} = K_{\mu\nu} + W_{\mu\nu}$ is the energy tensor,            
        $\Delta n(\xi_N)$ {and} $\rho (\xi_N)$ are the number of particles and particle density in a bin centred at $\xi_N$, {respectively, and} ${\Delta \Omega}_{ij} = \Delta \Omega(\xi_N, {\Delta \xi_N},\mathbf{r}_i^\alpha,\mathbf{r}_j^\beta)$ is the path (contour) joining positions of particles $i$ and $j$, which is determined by means of the Irving--Kirkwood contour (IKC)~\cite{irving1950statistical}. 
        This contour weights the contribution of each particle pair to the virial tensor based on the geometry of the interface and the orientation of the forces between particles relative to that interface, defining a straight path 
        \[\mathbf{r}(\lambda) 
            = \frac{1}{2}(\mathbf{r}_i^\alpha+\mathbf{r}_j^\beta) + \frac{\lambda}{2} (\mathbf{r}_j^\beta-\mathbf{r}_i^\alpha)\] joining the positions of all atom pairs, given a free parameter $-1 \leq \lambda \leq 1$. Calculations reported in this work use similar {IKC} parametrizations for planar interfaces as the ones shown in Refs.~\citenum{ikeshoji2003molecular} and \citenum{lu2022atomistic}, while for spherical interfaces the {IKC} is expressed as in Ref.~\cite{thompson1984molecular}.
            More details are provided in 
            the \ref{app:A}.
        \subsubsection{Interfacial tension}
            In this work, we explore how pressure tensors and the interfacial properties of He/{LLLA} mixtures (interfacial tension, interfacial density, adsorption, etc.) are affected by the composition of the {LMA}. 
            These properties are studied across selected states at a specific isotherm (1021.4~K) at which all possible lead--lithium alloys are molten. This is done by varying the ratio of lead to lithium atoms in the solvent.
            We pay special attention to changes {of interfacial phenomena induced by modifications of the atomic composition of the solvent.}
            Lead--lithium mixtures are known to be thermodynamically non-ideal~\cite{alvarez2025henry, buxbaum1984chemical}, a fact that 
            induces non-negligible mixing effects.

            \paragraph{Spherical interfaces}
            \begin{figure}[t]
                \centering
                \includegraphics[width=0.4\textwidth]{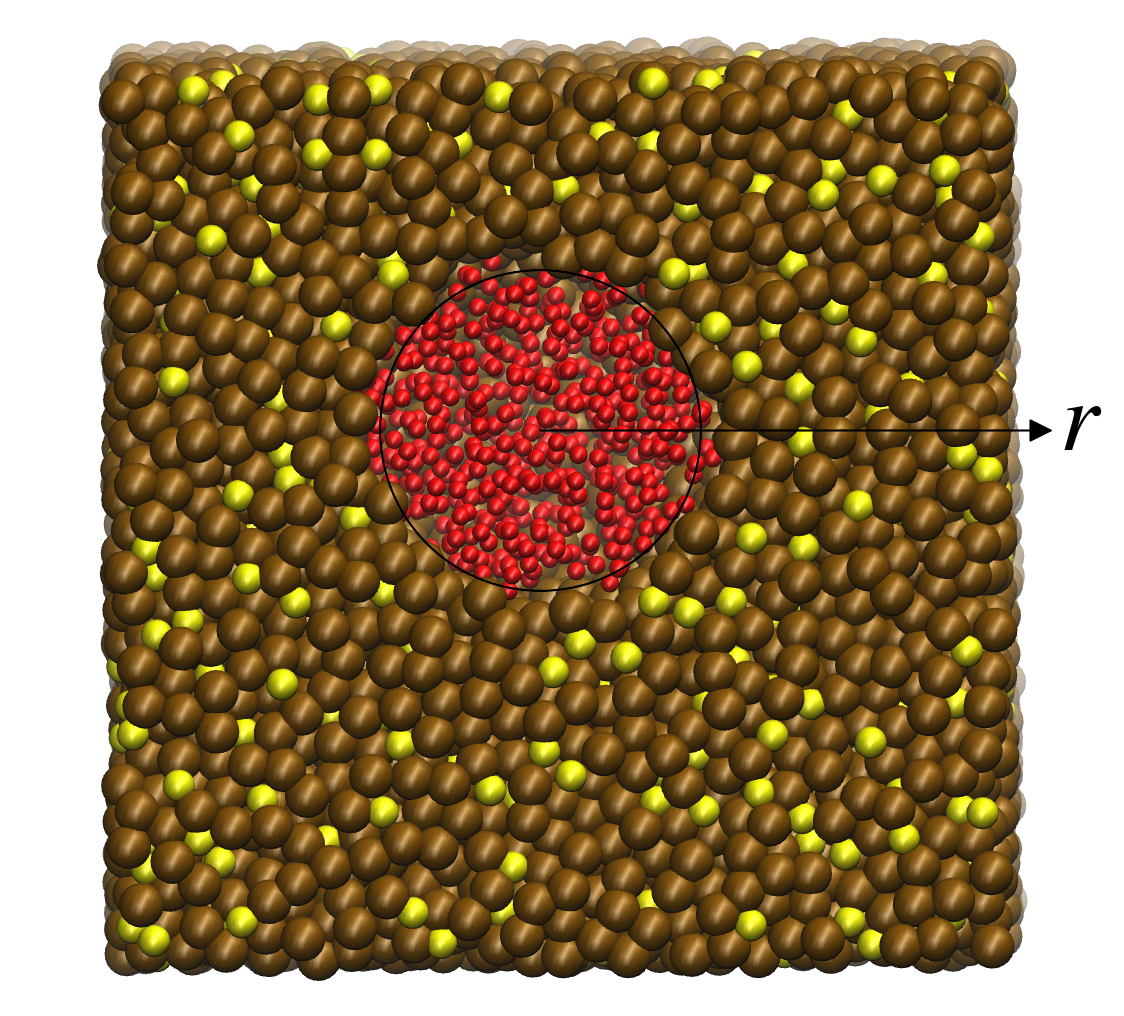}
                \caption{Snapshot of a helium (red) bubble within a eutectic lead--lithium (brown, yellow, respectively) mixture.
                }
                \label{fig:bubble}
            \end{figure}
            We consider isolated helium clusters embedded in a bulk liquid solution (see Fig.~\ref{fig:bubble} for a representative configuration). The system size is selected to ensure that interactions between the cluster of interest and any periodic replicas or additional clusters are negligible.
            Clusters are assumed to be (i) spherical, and (ii) in equilibrium with the surrounding {LM} phase.

            As a general fact, {the values of the interfacial tension depend on the curvature of the interface}, {which depends on} the choice of the arbitrary dividing surface of radius $\mathcal{R}$.
            Historically, two natural choices for its definition have been proposed \cite{thompson1984molecular, montero2019interfacial, montero2020interfacial}: $\mathcal{R}_e$ (equimolar radius) and $\mathcal{R}_s$ (radius of the surface of tension).
            The former defines a surface where the atom density in the clustered and bulk gases is the same, while the latter defines the surface at which $\gamma_s\equiv \gamma(\mathcal{R}_s)$ satisfies the Young--Laplace equation~\cite{young1805iii,thompson1984molecular}: 
            \begin{equation}\label{eq:YL}
                \displaystyle {(\Delta p)_s} = \frac{ 2 \gamma_s}{\mathcal{R}_s} \ ,
            \end{equation}
            where ${(\Delta p)_s} \equiv p_c - p_d > 0$ is the {difference
            between the pressure of} helium in the cluster and in the {LM} solution, and $\mathcal{R}_s$ {stands for} the radius of the spherical interface.
            {The definition of $\mathcal{R}_s$} is not arbitrary but, on the grounds of mechanical considerations\cite{thompson1984molecular}, it needs to be consistent with            
            \begin{eqnarray}\label{eq:radius_Thompson}
                \mathcal{R}_s = \left[\langle r^3 \rangle_s\right]^{1/3} \ ,
            \end{eqnarray}
            where $r \equiv |\mathbf{r} - \mathbf{r}_{CoM}|$.
            The discrepancy between the two former radii is typically quantified by the Tolman length~\cite{tolman1949effect, thompson1984molecular, montero2020interfacial}, $\displaystyle \delta_T \equiv \lim_{\mathcal{R}_s \rightarrow \infty} [\mathcal{R}_e - \mathcal{R}_s] $.

            Here, let us introduce the shorthanded notation           \begin{eqnarray}\label{eq:moms_sph}
                \langle r^n \rangle_s &=& \int_0^\infty r^n f_s(r) \ \mathrm{d}r
                \ ,
                \label{nmom_s}
                \\
                f_s(r) &\equiv& -\frac{1}{(\Delta p)_s} \frac{\mathrm{d}p_N(r)}{\mathrm{d}r} 
                \nonumber\\
                &=& + \frac{1}{(\Delta p)_s} \frac{2}{r} \left[p_N(r) - p_T(r)\right] \ . 
            \end{eqnarray}
            
            The pressure difference between the phases at each sides of spherical interfaces (inner--outer) then is         \begin{eqnarray}\label{eq:difpress_Thompson}
                (\Delta p)_s = -\int_0^\infty \frac{\mathrm{d}p_N(r)}{\mathrm{d}r}\mathrm{d}r > 0 \ ,
                \nonumber
            \end{eqnarray}
            which normalizes Eq.~(\ref{nmom_s}) (0th-moment):
            \begin{eqnarray}
                1 = \langle r^0 \rangle_s = \int_0^\infty f_s(r) \ \mathrm{d}r \ .
                \nonumber
            \end{eqnarray}
            
            Then, from Eq.~(\ref{eq:radius_Thompson}), the cube of $\mathcal{R}_s$ is understood as the $3rd$-moment of the radial coordinate $r$, given a weighting function $f_s(r)$.
            This is, in essence, the method proposed by Thompson et al.~\cite{thompson1984molecular}. Then, by substituting Eq.~(\ref{eq:radius_Thompson}) into the {YL} equation (\ref{eq:YL}), it follows that:
            \begin{eqnarray}\label{gamma}
                \gamma_s = 
                \left[\frac{1}{8} {(\Delta p)_s}^3 \int_0^\infty r^3 f_s(r) \ \mathrm{d}r \right]^{1/3} \ .
            \end{eqnarray}
            
            More generally, quantity $\langle r^n \rangle_s$ evaluates the cumulative contribution of the $n$-th powers of $r$ starting from the centre of the spherical phase up to long distances ($r \rightarrow \infty$, where interfacial inhomogeneities vanish), weighted by the local differences between normal and tangential pressure components:
            \[
                \Delta p (r) \equiv p_N(r)-p_T(r) \ , 
            \]
            where
            \[\begin{cases}
                p_N(r) = \Pi_{rr}(r) \ ,
                \\
                \displaystyle p_T(r) = \frac{\Pi_{\theta\theta}(r) + \Pi_{\phi\phi}(r)}{2} \ .
            \end{cases}        
            \]
            This formulation naturally accounts for interfacial inhomogeneities, since regions where the pressure tensor deviates from isotropy contribute most significantly to the integral in Eq.(\ref{gamma}). As a result, $\langle r^n \rangle_s$ provides a compact measure of how microscopic structural variations across the interface influence macroscopic interfacial properties, linking local pressure anisotropies to global thermodynamic observables.

            \paragraph{Planar interfaces}
            \begin{figure}[t]
                \centering
                \includegraphics[width=0.5\textwidth]{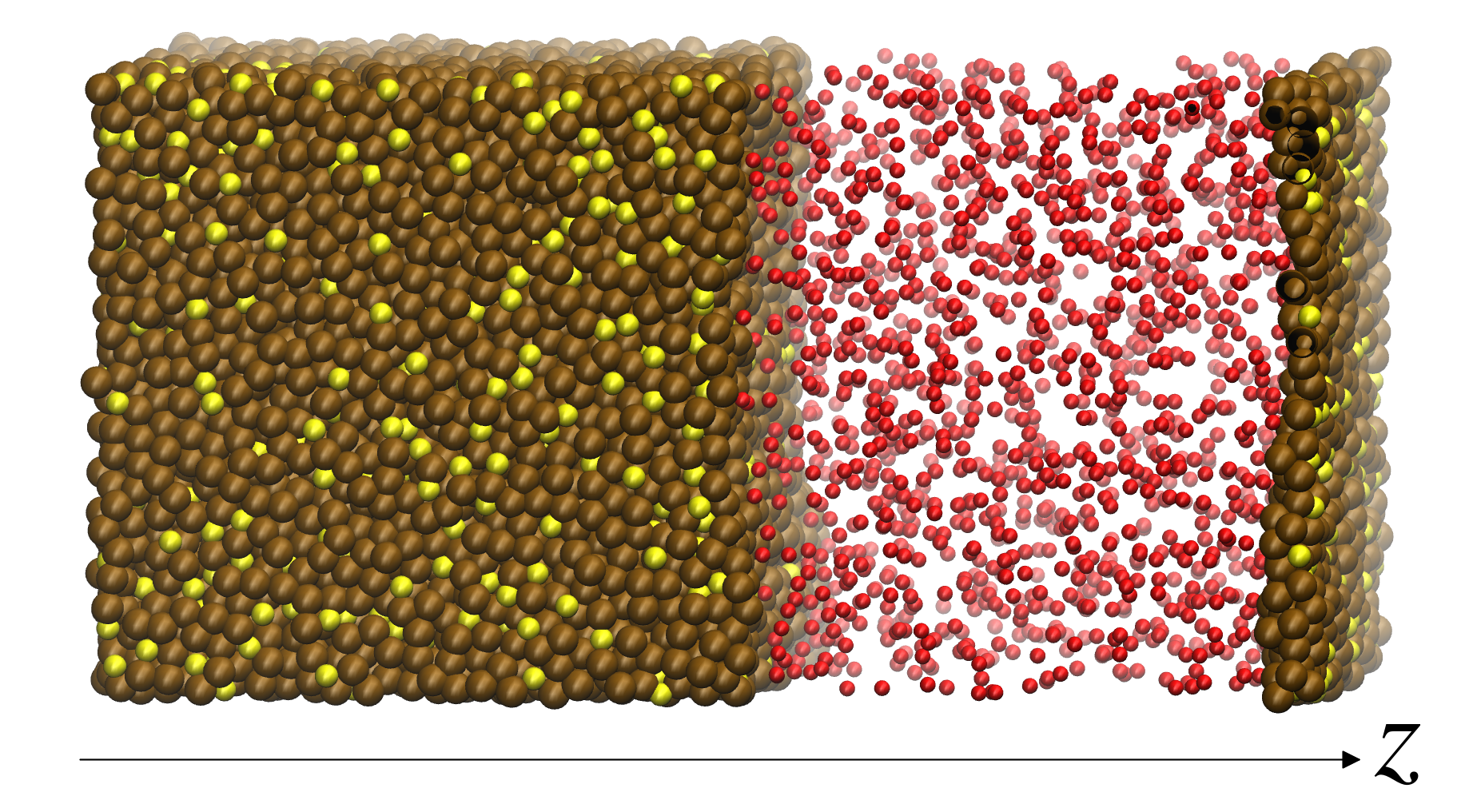}
                \caption{Snapshot of a He/LLE double layer, appearing after imposing PBC. Same colour coding as in Fig.~\ref{fig:bubble}.}
                \label{fig:bilayer}
            \end{figure}
            Planar symmetries stand as the limiting case for spherical interfaces with infinite {radius.} 
            Given a flat interface between the gas and liquid phases, whose direction is normal to the $z$-axis, we get:
            \[
            \begin{cases}
                        \displaystyle f_p(z) = \frac{1}{L_z (\Delta p)_p} \left[p_N(z) - p_T(z) \right] \ ,
                        \\
                        \Delta p (z) \equiv p_N(z) - p_T(z) \ .
            \end{cases}
            \]
            Here, $L_z$ is the total length of a simulation box along the direction perpendicular to the flat interface, while normal and tangential components of pressure tensors respectively read as:
            \[
                \begin{cases}
                    p_N(z) = \Pi_{zz} (z) \ ,
                    \\
                    \displaystyle p_T(z) = \frac{\Pi_{xx}(z) + \Pi_{yy}(z)}{2} \ .
                \end{cases}
            \]
            Hence, $n$ moments now are evaluated along the whole $z$-axis,
            \begin{eqnarray}\label{nmom_p}
                \langle z^n \rangle_p &=& \int_{-\infty}^\infty z^n f_p(z)\ \mathrm{d}z 
                \nonumber\\
                &\simeq& \int_{-L_z/2}^{L_z/2} z^n f_p(z)\ \mathrm{d}z \ .
            \end{eqnarray}
            In passing from the infinite integration limits to the {finite size of the 
            simulation boxes}, we assumed that {the inhomogeneities in the pressure} vanish far from the interface, namely $\Delta p(z) \simeq 0$ for both $z \ll z_0$ and $z \gg z_0$, where $z_0$ indicates the location of the interface.

            When a single interface is considered along the whole range of $z$ ($N_p=1$), and the coordinate system is chosen to have its origin at the position {of the} interface, the expected value of $z$ is $0 \equiv z_p = \langle z \rangle_p$. 
            In that case, $f_p(z)$ profiles are bell-shaped {functions,  $b(z)$}, and vanish far from the interface region. 
            Several consecutive interfaces ($N_p>1$) may be considered by joining several equally-sized boxes, having a total simulation length of $L_z' = N_p L_z$.
            Without loss of generality, the expected value of $z$ reads:
            \begin{eqnarray}
                \langle z \rangle_p &=& \frac{1}{L_z'}\int_{-L_z/2}^{L_z/2} z \ f_p(z) \  \mathrm{d}z
                \nonumber\\
                &=& \frac{1}{N_p  L_z}\int_{-L_z/2}^{L_z/2} z \ \dfrac{\Delta p(z)}{(\Delta p)_p} \  \mathrm{d}z
                \nonumber\\
                &=& \dfrac{1}{N_p} \sum_{s=1}^{N_p} z_s \ .
            \end{eqnarray}
            Now, the weighting function is decomposed into several bell-shaped functions, $f_p(z) = \sum_{s=1}^{N_p} b(z-z_s)$.
            Also notice that $f_p(z)$ would not be normalized to the unity but to the total number of interfaces, \[\int_{-\infty}^{\infty} f_p(z) \ \mathrm{d}z = N_p \ .\]
            Fig.~\ref{fig:bilayer} shows a representative He/Pb16Li bilayer (i.e., $N_p=2$), corresponding to the minimal number of interfaces that appear under PBC.

            This formalism yields, in essence, to well-known expression for the planar interfacial tension between planar obtained from the integral of the pressure anisotropy profile~\cite{kirkwood1949statistical, yang2014diffusion, lu2022atomistic}:
            \begin{equation}
                \gamma_p = \frac{1}{N_p}\int_{-\infty}^\infty {\Delta p}(z) \ \mathrm{d}z  \ .
            \end{equation}

        \subsubsection{Ideal behaviour of pressure tensors {in spherical geometries}}
            \begin{figure}[t]
                \centering
                \includegraphics[width=0.45\textwidth]{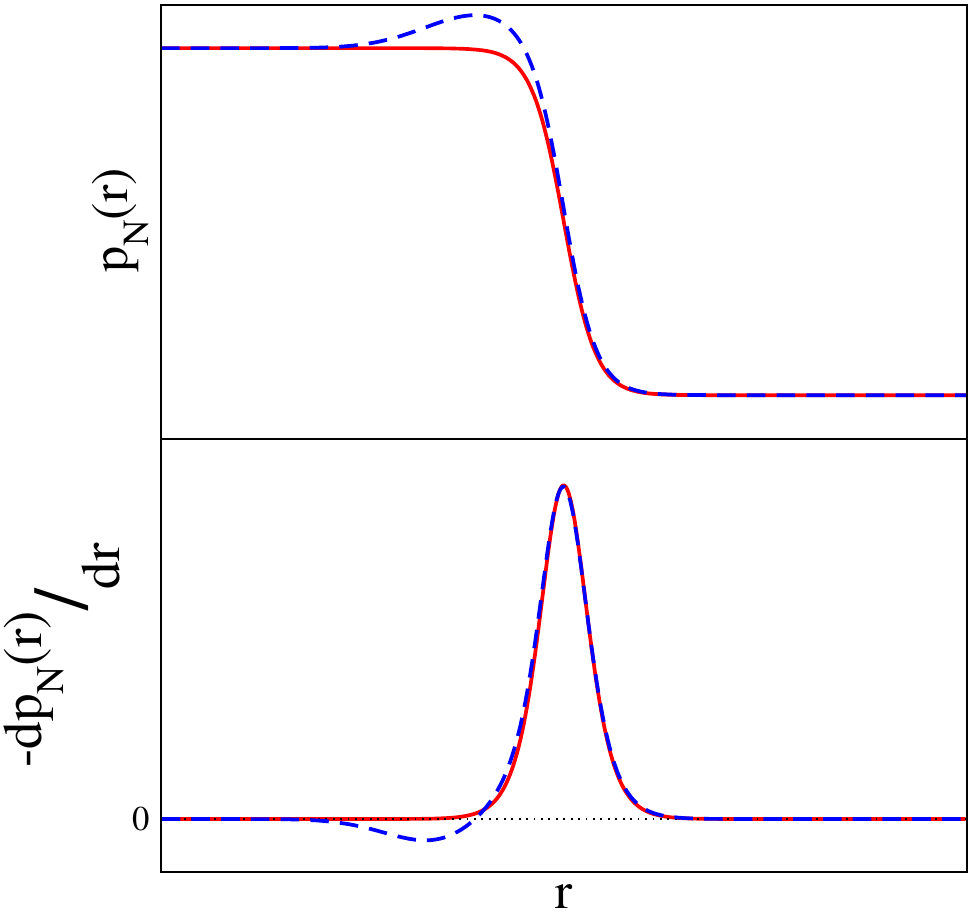}
                \caption{Ideal (red solid line) and non-ideal (blue dashed line) radial normal components of the pressure tensor (top panel) and the opposite of its derivative (bottom panel). 
                }
                \label{fig:pN_ideal}
            \end{figure}
            In the case of spherical interfaces, it seems convenient to remark that normal profiles are expected to obey the following ideal expression~\cite{thompson1984molecular}:
            \begin{eqnarray}\label{eq:pN_ideal}
                p_N^\text{id} (r) && = \frac{1}{2} \left[ p_N(0) + p_N(\infty) \right] \nonumber\\
                &&- \frac{1}{2} \left[ p_N(0) - p_N(\infty) \right] \nonumber\\ &&\times\tanh\left(\frac{2 (r-r_{mid})}{r_{w}}\right)
                \ ,
                \nonumber
            \end{eqnarray}
            where $r_{mid}$ and $r_{w}$ {are} the midpoint and the width of the interface, respectively, {whereas $p_N(0)$ and $p_N(\infty)$ are the normal pressure tensor at zero and infinite distances, respectively}.
            Then, the weighting function introduced in Eq.(\ref{eq:moms_sph}) {reads as:}
            \begin{eqnarray}
                f_s^\text{id}(r) &=& - \frac{1}{[p_N(0)-p_N(\infty)]} \dfrac{\mathrm{d}p_N^\text{id}}{\mathrm{d}r} \nonumber\\
                &=& \frac{1}{r_w} \left[1-\tanh^2{\left(\dfrac{2(r-r_w)}{r_w}\right)}\right] \nonumber\\
                &\geq& 0 .
                \nonumber
            \end{eqnarray}
            In other words, when considering radial paths centred at helium bubbles, the pressure is expected to decrease monotonically from the uniform pressure inside the bubble to the lower bulk pressure corresponding to the surrounding {LM}. Then, $f_s^\text{id}(r)$ can be formally treated as a density probability function. The generalizability of this statement when {LMA}s behave {as} non-ideal mixtures will be discussed below in Section~\ref{sec:RESULTS}.
            
            In Figure~\ref{fig:pN_ideal}, 
            $p_N^\text{id}(r)$ and its derivative
            $-\dfrac{\mathrm{d}p_N^\text{id}}{\mathrm{d}r} =${$f_s^\text{id}(r)$} $\Delta p$, where
            \[{\Delta p \equiv p_N(0)-p_N(\infty)} \ ,\] are represented.
            We can observe the qualitative trend for the deviation of the profiles in the non-ideal case. 
            
            In Fig.~\ref{fig:pN_ideal} qualitative trends of $p_N(r)$ , which include (i) the asymptotic approach to zero of both ideal and non-ideal profiles and, (ii) the fact that the derivative is negatively defined for all $r$ in the ideal case, while the non-ideal derivative shows a small positive region.
            Possible departures from the ideal case, due to local increments of $p_N$ near the interface, are depicted with blue dashed lines, yielding to finite regions at which $f_s^\text{id}(r)$ is allowed to show negative values.
            {Despite} there is no thermodynamic requirement  enforcing $p_N(r)$ to be maximum at $r=0$,  it can increase near the location of the interface.

            In any case, we expect that the normal pressure tensor
            should asymptotically reach two limit values: \[p_N(r) \simeq \begin{cases}
                p_N(0)\ \ , \ \ \forall r \ll r_0
                \\
                p_N(\infty), \ \ \forall r \gg r_\infty
            \end{cases} , \] where $r_0$ and $r_\infty$ {are} the arbitrary thresholds {indicating the distances below and above} which normal--tangential pressure differences vanish ($\Delta p = 0$), respectively.
            We should also notice that, despite the local allowance of $-\dfrac{\mathrm{d}p_N}{\mathrm{d}r} < 0$, the overall pressure difference must be strictly positive,
            \[
                (\Delta p)_s \equiv p_N(0) - p_N(\infty) = - \int_0^\infty \dfrac{\mathrm{d}p_N}{\mathrm{d}r} \ \mathrm{d}r > 0 ,
            \]
            in order to ensure the stability of the spherical cluster and in good agreement with the general theory of surface tension theory in spherical systems~\cite{gibbs1878equilibrium, gibbs1906scientific, young1805iii, buff1951spherical, buff1955spherical,montero2020young}.
            
\section{\label{sec:RESULTS}Results and Discussion}
    Evaluating the accuracy of the values of interfacial tension from the simulations is challenging due to the lack of reliable experimental characterization for He/Li, He/Pb and He/LLE interfaces. However, the values presented in this work are not meaningless. The following benchmarks can serve as useful reference points:
    \begin{enumerate}
        \item The knowledge of experimental values of the surface tension of bulk lithium\cite{davison1968compilation,zinkle1998summary}, around 0.3--0.4 N/m;
        \item  The experimental values of the surface tension of bulk {LLE}~\cite{de2008lead};
        \item {The} experimental surface tension of ultracold helium, which ranges between 0.35 N/m at ultra-low temperatures around 1K, down to 0.1 N/m at the boiling point around 4~K\cite{wohlfarth2008surface}. In this case, no data has been found at higher temperatures, but one clearly expects surface tension values to drop for increasing temperatures. 
    \end{enumerate}
    These three data sets are considered for comparative purposes only because, despite sharing the same dimensionality, surface tension and interfacial tension represent distinct physical phenomena. The former generally refers to the property of a liquid surface in contact with a gas (usually air) or a vacuum, while the latter describes the tension existing at the boundary between any two immiscible or partially miscible condensed phases, such as two liquids or a liquid and a solid.

    \subsection{Pressure tensors and interfacial tension at He/LLE, He/Li and He/Pb interfaces}
        As a starting point, we study the convergence of values of interfacial tension as the size of the helium cluster increases. Let us consider the paradigmatic case of helium nanobubbles within the eutectic lead--lithium alloy (16\%~Li--84\%~Pb at $T=508$~K).

        \begin{figure}[!t]
            \centering
            \includegraphics[width=0.5\textwidth]{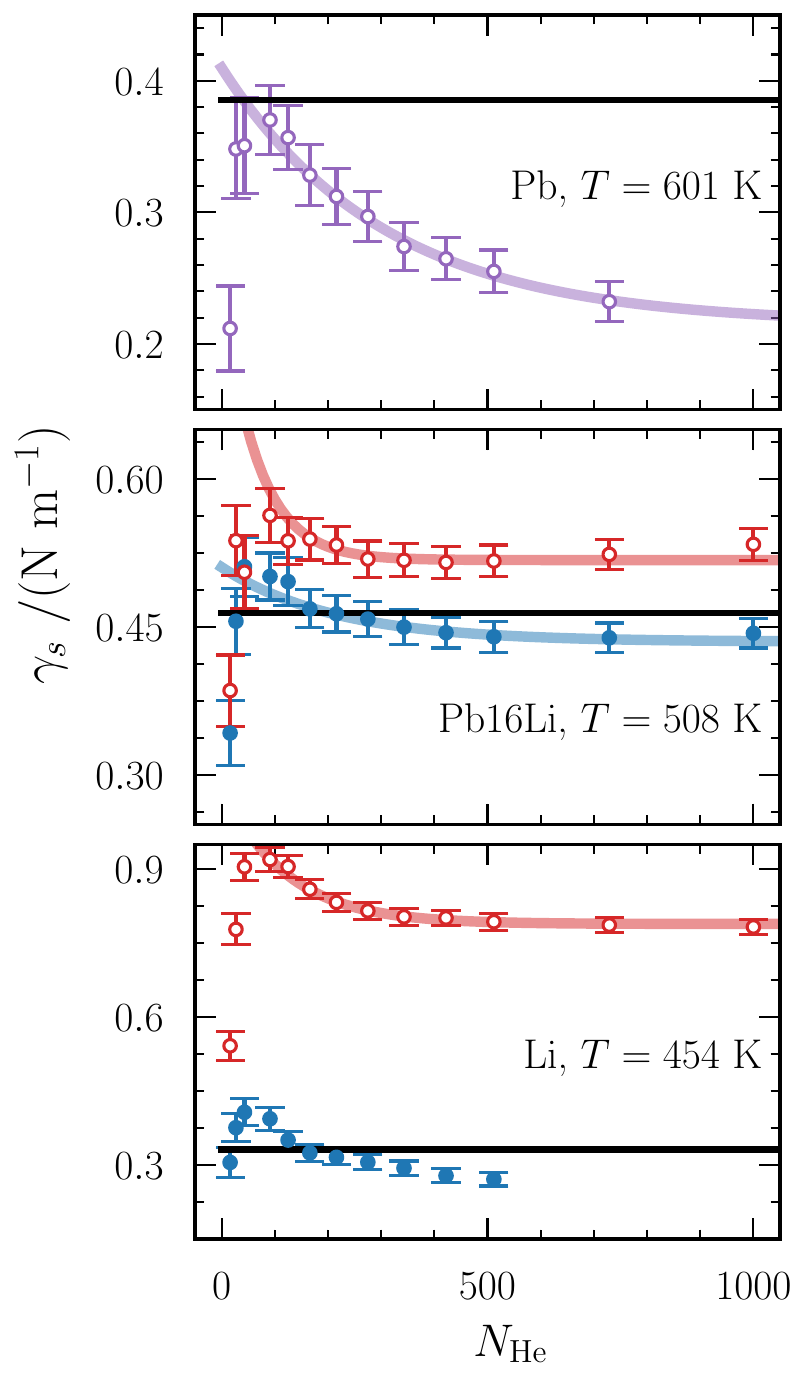}
            \caption{Interfacial tension values for helium bubbles in the 3 indicated {LM}s, given a representative temperature for each one.
            Black lines represent experimental surface tension values for each {LM} at same temperatures (Pb--Passerone et al.\cite{passerone1983surface}, Pb16Li--Buxbaum\cite{buxbaum1984chemical}, and Li--Ohse\cite{ohse1985handbook}). 
            \\
            Top panel: Violet circles represent correspond to simulations of helium bubbles in pure Belashchenko-like lead\cite{belashchenko2012computer,belashchenko2019inclusion}. 
            \\
            Middle panel: Blue dots and red circles represent interfacial tension values between helium cluster (made up by $N_{\rm He}$ atoms) and Pb16Li at 508~K (namely, the {LLE} alloy), where the latter has been parametrized as in Refs.~\cite{al2023parametrization} and \cite{belashchenko2019inclusion} (respectively).
            \\
            Bottom panel:
            Analogously, blue dots and red circles represent simulations for He/Li spherical interfaces using the Al-Awad et al.~\cite{al2023parametrization} and Belashchenko~\cite{belashchenko2019inclusion} EAM potentials for lithium.}            
            \label{fig:convergence_tri_mp}
        \end{figure}

        In Figure~\ref{fig:convergence_tri_mp}, we compare the interfacial tension of helium spherical clusters within LLE and in pure lithium and lead solvents, using two different parametrizations of the LM--EAM potential (Al--Awad et al.~\cite{al2023parametrization} and Belashchenko~\cite{belashchenko2019inclusion}). 
        We can observe that, in the LLE, $\gamma_s$ rapidly converges to constant values. In particular, one can realize that there is an adequate matching between (a) the interfacial tension from simulations of large cluster--{LLE} (Al-Awad's model for Pb16Li\cite{al2023parametrization}, at 508~K), and (b) the experimental surface tension values of the eutectic alloy at the same temperature~\cite{de2008lead}.
        This agreement not only further corroborates the enhanced performance of Al-Awad's parametrization for the lead--lithium alloy potential, but also indicates that helium clusters behave as bubbles rather than droplets. Consequently, the interfacial tension between the two immiscible phases is effectively identical to the surface tension of the liquid metal in contact with a gas or vacuum.
        Further, we can also point out that the parameterization of Al-Awad et al. of the EAM alloy potential is particularly suitable to represent the eutectic mixture, while for pure lead at its normal melting point interfacial tension deviates from its surface tension value in a quite large extent, showing how subtle are the differences in suitability for EAM models depending on the composition of the LM.

        \begin{figure}[!t]
            \centering
            \includegraphics[width=0.5\textwidth]{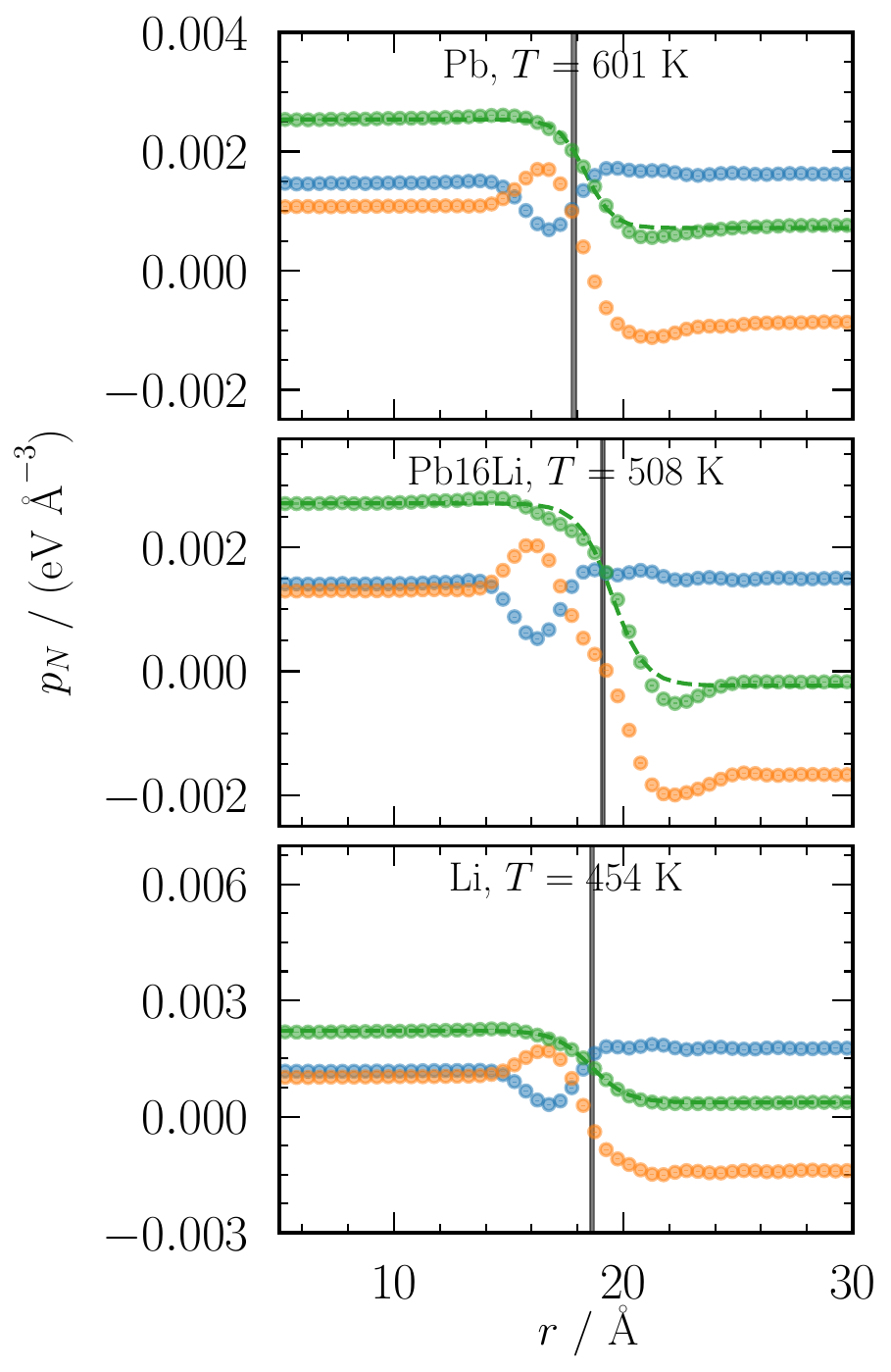}
            \caption{Kinetic (blue), configurational (orange) and total (green) radial normal pressure tensor components for interfaces between helium spherical clusters of $N_{\rm He}=512$) and Al-Awad--EAM liquid metals~\cite{al2023parametrization} above their normal melting point temperatures: (a) Li at 454~K; (b) Pb at 601~K; and, (c) Pb16Li at 508~K.
            In addition, interfacial radii ($\mathcal{R}_s$) obtained after integrating the corresponding normalized profiles have been depicted as gray vertical lines, whose width indicate their uncertainty. } 
            \label{fig:radial_stress_above_mp_A}
        \end{figure}
        \begin{figure}[!t]
            \centering
            \includegraphics[width=0.5\textwidth]{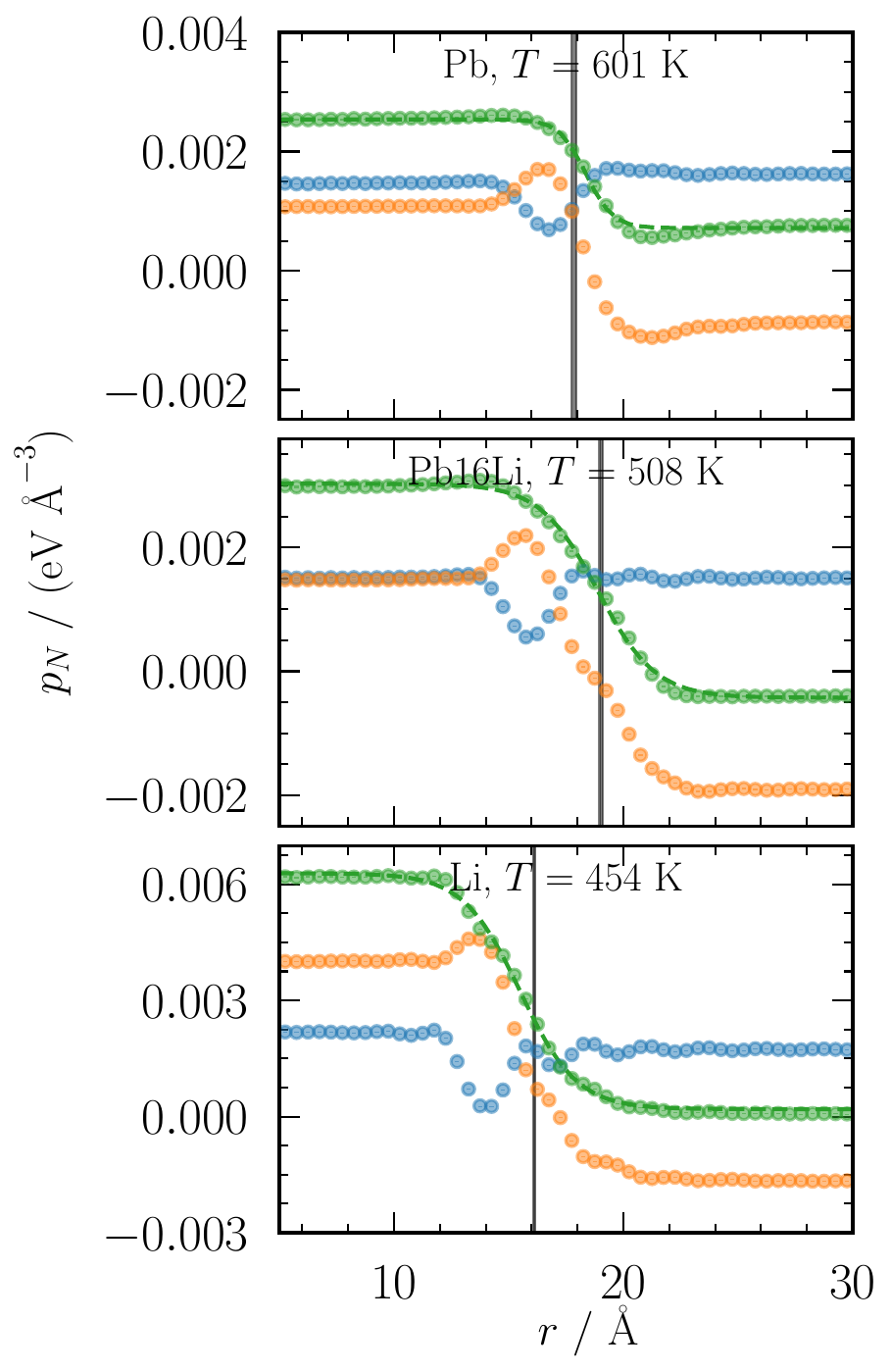}
            \caption{Kinetic (blue), configurational (orange) and total (green) radial normal pressure tensor components for interfaces between helium spherical clusters of $N_{\rm He}=512$) and Belashchenko--EAM liquid metals~\cite{belashchenko2019inclusion} above their normal melting point temperatures: (a) Li at 454~K; (b) Pb at 601~K; and, (c) Pb16Li at 508~K.
            As in Fig.~\ref{fig:radial_stress_above_mp_A}, vertical lines indicate the distance between the position of the interface and to the centre of the bubble (namely the radius of the interface).}
            \label{fig:radial_stress_above_mp_B}
        \end{figure}
        As representative cases, radial pressure tensors, with origin in the {CoM} 
        of helium clusters of $N_{\rm He}=512$ atoms, are displayed in Figures~\ref{fig:radial_stress_above_mp_A} and \ref{fig:radial_stress_above_mp_B}. 
        Each panel shows kinetic{, 
        \[
            p_k(r) = \beta^{-1} \left\langle\rho (r)\right\rangle 
            \ ,
        \]}
        configurational{, 
        \begin{equation}\label{eq:pVr}
            p_v(r) = \frac{1}{2}\left\langle \sum_{\alpha, i} \sum_{\beta, j} [\mathbf{r}_{ij}^{\alpha\beta}]_\mu [\mathbf{f}_{ij}^{\alpha\beta}]_\nu \  \frac{\partial \Omega_{ij}^{\alpha\beta}}{\partial r} \right\rangle 
            \ ,
        \end{equation}}
        and total,
        \[
            p_N(r) = p_k(r) + p_v(r) \ ,
        \]
        terms, evaluated above the normal melting points of each solvent. 
        In Equation~\ref{eq:pVr}, the IKC can be expressed, {in good agreement with} Ref.~\cite{thompson1984molecular}, as:
        \begin{eqnarray}
            &&\left.\frac{\partial\Omega_{ij}^{\alpha\beta}}{\partial r}\right|_s
            = \lim_{\Delta r \rightarrow 0} \frac{\Delta \Omega_s(r,{\Delta r},r_i,r_j)}{\Delta r}
            \nonumber\\
            &&= \frac{1}{r} 
            \left[ \theta(|\lambda_+|-1) + \theta(|\lambda_-|-1) \right.
            \nonumber\\
            && \ \ \ \  - \left. \theta(|\lambda_+|-1)\times\theta(|\lambda_-|-1) \right. \nonumber\\
            && \ \ \ \  \left.+ 2\times\theta(1-|\lambda_+|)\times\theta(1-|\lambda_-|) \right]
            \ , \ \ 
            \nonumber
        \end{eqnarray}
        where $\theta (x)$ denotes the Heaviside step function.
        Figures~\ref{fig:radial_stress_above_mp_A} and \ref{fig:radial_stress_above_mp_B} reveal depletion of atoms at the vicinity of interfaces (see vertical lines). Consequent reductions in $p_k$ correlate with increments in $p_v$. Here, the interpretation is clear: the strengthening of helium near interfaces empties that region, thus reducing the local kinetic activity.
        Moreover, after summing the two contributions, one can realize that in the three reported states the two corresponding anomalies compensate, leading to non-increasing functions of the radial distance, roughly represented by Equation~\ref{eq:pN_ideal}.
    \subsection{{Effects of concentration of lithium on the surface tension of He bubbles along the 1021 K isotherm}}
        \begin{figure}[!t]
            \centering
            \includegraphics[width=0.5\textwidth]{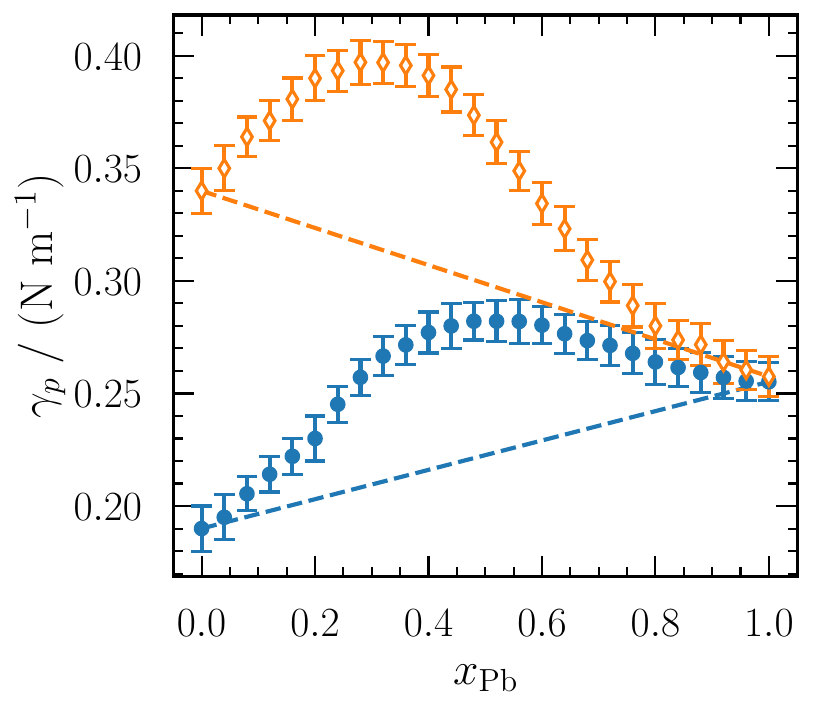}
            \caption{{Interfacial tension of a planar surface} as function of the solvent atomic fraction (composition) along the 1021~K isotherm. 
            Blue and orange circles correspond to {IT} values where lead--lithium alloys were modelled by AEAM-2023\cite{al2023parametrization} and BEAM-2019 \cite{belashchenko2019inclusion} potentials, respectively.
            Both cases consider {TTS}\cite{sheng2021development} and {SBMILL}\cite{sladek2014ab} potentials for helium interactions.
            Dashed lines indicate ideal-mixture predictions.}
            \label{fig:gamma_vs_x}
        \end{figure}
\begin{figure}[!t]
            \centering
            \includegraphics[width=0.5\textwidth]{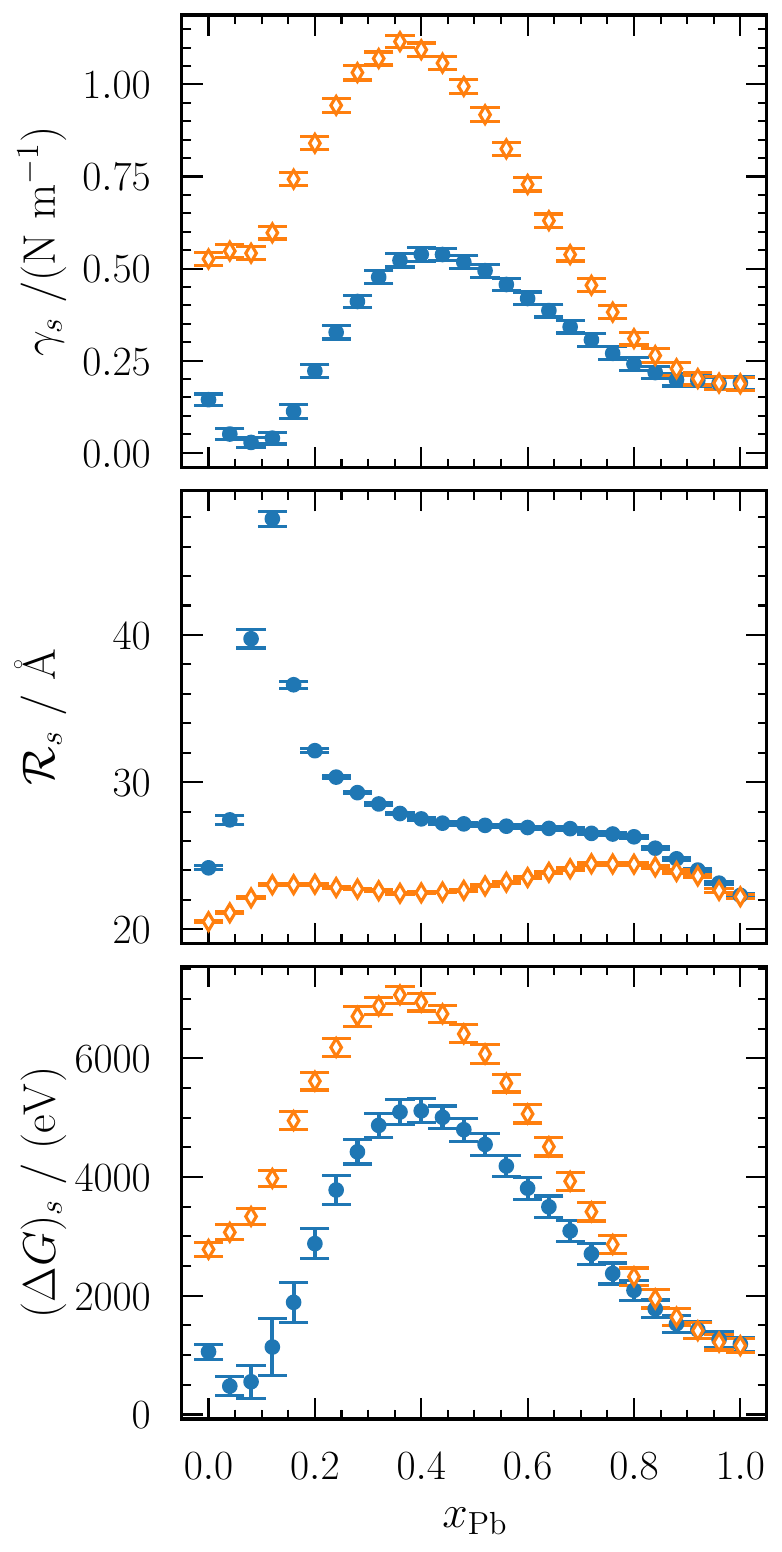}
            \caption{{Interfacial tension (top), interfacial radii (middle), and interfacial free energy variations (bottom), as function of the solvent composition ($x_{\rm Pb} = 1 - x_{\rm Li}$). 
            }
            All cases consider He/Pb--Li spherical interfaces, where all helium atoms segregated into a single spherical cluster of $N_{\rm He}=512$. Blue circles and orange (empty) diamonds indicate that the {LM} phase has been modelled with the Al-Awad's\cite{al2023parametrization} or Belashchenko's\cite{belashchenko2019inclusion} (respectively){, combined with the same potentials for helium atoms ({TTS} for He--He and Li--He~\cite{sheng2020conformal,sheng2021development}, {SBMILL} for Pb--He)}.}
            \label{fig:tension_spherical}
        \end{figure}

        After the analysis of He/LM interfacial tension at the technologically interesting states, we analyse analogue values in the full range of lead concentrations ($x_{\rm Pb} \equiv 1 - x_{\rm Li}$) in the LM. In particular, we report the values of $\gamma_p$ as a function of $x_{\rm Pb}$ in Figure~\ref{fig:gamma_vs_x} for a planar interface, where we compare our results using the models of Al-Awad et al.~\cite{al2023parametrization} and Belashchenko~\cite{belashchenko2019inclusion} with the expected values of $\gamma_p$ in the case of considering an ideal mixture. We can investigate the non-idealities observed in Figs.~\ref{fig:gamma_vs_x} and~\ref{fig:tension_spherical} by 
        drawing on some physical insights. In this way, non-ideal mixture effects of the {LM} alloys were previously shown to be important with regard to solubility of helium atoms~\cite{alvarez2025henry}. As the most relevant trend, in Figure~\ref{fig:gamma_vs_x} we see that $\gamma_p$ is maximized at some intermediate composition, regardless of the chosen potentials for the lead--lithium alloy. So, in planar cases ($\mathcal{R}_s \rightarrow \infty$) curvature vanishes homogeneously for all {LM} compositions, and changes on interfacial tension values are considerably reduced.
        Even so, a maximum appears in each set of calculations: In the case of Al-Awad's Pb--Li\cite{al2023parametrization}, interfacial tension maximizes when the amounts of the two metals are similar ($x_{\rm Li} \simeq x_{\rm Pb} \simeq 0.5$)
        whereas in the case of Belashchenko's Pb--Li\cite{belashchenko2019inclusion}, interfacial tension reaches its maximum value for $x_{\rm Pb} \simeq 0.3$ ($x_{\rm Li} \simeq 0.7$).

        In Figure~\ref{fig:tension_spherical} we can corroborate that this intermediate-composition mixture effect is not only present but even more accentuated for spherical interfaces. There, we report a comparison between interfacial tension, the radius of the He/LLA interface ($\mathcal{R}_s = \sqrt[3]{\langle r^3 \rangle_s}$) and the {interfacial} free energy difference.
        As it was pointed out in Ref.~\cite{alvarez2023nucleation}, the last quantity can be computed as
        \begin{eqnarray}
                (\Delta G)_{s} &=& W(\mathcal{R}_s) = 4 \pi {\mathcal{R}_s}^2 \ \gamma_s(\mathcal{R}_s) 
                \nonumber
                \\ 
                &=& -2 \pi \int_0^\infty r^3 \frac{\mathrm{d}p_N(r)}{\mathrm{d}r}\mathrm{d}r > 0 \ .
        \end{eqnarray}

        In planar symmetries, interfacial tension is maximized at atomic fractions where the shell of first-neighbours contains more lithium atoms. As lithium--helium repulsions become stronger, interfacial tension {reaches its maximum} when lithium atoms are most concentrated at interfaces.
    
        In the case of spherical symmetries, in addition to the aforementioned effect due to the arrangement of solvent atoms at the vicinities of interfaces, curvature also appears as a crucial factor.
        In Figure~\ref{fig:tension_spherical} we show that the values of interfacial tension maximize at the following alloy compositions: (1) $44\%\;\text{Pb},\ 56\%\;\text{Li}$ when the alloy model is parametrized as in Ref.~\cite{al2023parametrization}; and (2) $40\%\;\text{Pb},\ 60\%\;\text{Li}$ when the alloy model is parametrized as in Ref.~\cite{belashchenko2019inclusion}. 

        Figures \ref{fig:pNr_512_A} and \ref{fig:pNr_512_B} of the SI display the characteristic pressure tensors for each state shown in Figure~\ref{fig:tension_spherical}. 
        In all cases, the positions of maxima in $p_v(r)$ and minima in $p_k(r)$ match consistently.
        This correlation indicates that the localized peak in forces effectively depletes particles from that specific radial distance from the bubble centre.
        Moreover, this position shifts to larger distances as the size of the bubbles increase.
    Interestingly, the sum of the two contributions --i.e., the total normal component of the pressure tensor profiles $p_N(r)$-- flattens for several compositions of the LM, from a short range of Li-rich alloys ({$0 \leq x_\text{Pb} = 1 - x_\text{Li} \lesssim 0.1$}) to a wider range of Pb-rich alloys ($0.75 \lesssim x_\text{Pb} \leq 1$), including the eutectic composition. In other words, the two contributions --namely, the depletion of atoms and the local increase of stress forces near the interface-- compensate and behave as the ideal profile $p_N^\text{id} (r)$, Eq.~\ref{eq:pN_ideal}.

    For intermediate ranges ({$0.1 \lesssim x_\text{Pb} = 1 - x_\text{Li} \lesssim 0.75$}), where the solubility considerably decreases and interfacial repulsions increase, the kinetic and configurational parts tend to lack balance, yielding to a local increase of the total pressure tensor.
    Generally, this `defect' appears at distances slightly larger than those where both the kinetic and virial terms contribute.
    As it has been shown in Figure \ref{fig:pN_ideal}, this local {loss of balance} is translated into an increase of $p_N(r)$ at the interface:
    \[
        p_N(r) - p_N^\text{id}(r) = \begin{cases}
            \neq 0, \ \ r_i < r < r_o
            \\
            0, \ \text{otherwise}
        \end{cases}\ ;
    \]
    where $r_i, r_o$ denote the inner and outer boundaries of the interface, respectively.
    In addition, while in the planar limiting case the location of the interfacial tension peak is found near an equimolar composition of the lead--lithium alloy, for curved interfaces it is shifted to mixtures richer in lithium.
    Consequently, in the `ideal' cases of compensating minimum kinetic and maximum virial contributions the normal pressure is found to be maximum inside the bubble, whereas in the `non-ideal' cases the normal pressure maximizes within the interface.

    Our analyses of pressure tensor suggest that curvature effects become more prominent when the normal component of the pressure tensors {increases locally with} respect to the ideal behaviour (see Eq.~\ref{eq:pN_ideal}).
    
    Radii corresponding to the analysed interfacial tension values, $\mathcal{R}_s$, are depicted in the bottom panel of Fig.~\ref{fig:tension_spherical}. From this plot, 
    one can observe an exaggerated expansion of helium bubbles in the AEAM-2023 $88\%\text{Li}-12\%\text{Pb}$ alloy, which could explain the local reduction in interfacial tension values in the corresponding region. The last observation points out the relevance of the curvature in the determination of the interfacial tension for spherical interfaces. This is a well-known feature, described by the Tolman expansion~\cite{tolman1949effect, thompson1984molecular, montero2019interfacial, montero2020interfacial}:
    \begin{equation}\label{eq:Tolman}
        \gamma_s = \gamma_\infty \left[1 - \dfrac{2 \delta_T}{\mathcal{R}_s} + \dots\right] \ ,
    \end{equation}
    where $\delta_T$ denotes the Tolman length.
    
    Finally, the bottom panel of Fig.~\ref{fig:tension_spherical} presents the Gibbs free energy differences between the clustered and corresponding metastable phases, i.e., the work required to form the spherical dividing interface. Consistent with the interfacial tension values, these profiles indicate that the energetic cost involved in the formation of such interfaces is maximized at a solvent composition of approximately 60\% Li--40\% Pb.
    The values of $(\Delta G)_s$ typically lie in the range $10^3-10^4$~eV, indicating that the associated free energy barriers are on the order of keV.

\section{\label{sec:CONCLUSIONS}Conclusions}
   
    In this work a detailed study of interfacial tensions and pressure tensors has been applied to helium--lead--lithium mixtures. Due to the rapid excess of the supersaturation threshold, noble gas atoms segregate from the liquid solution. Understanding this feature is crucial for the successful design of the future generation of nuclear fusion plants.
    The possible formation of helium bubbles in breeding is a concerning factor that needs to be quantified in order to enhance operational performance and safety.
    
    By using classical molecular dynamics, we provide an atomistic perspective on these complex mixtures, assessing a problem where practical technological requirements go hand in hand with theoretical fundamental scenarios.
    The modelling of the system relies on the choice for the interatomic potentials.
    Because of the nature of the proposed force field, helium tends to segregate and form one or several clusters.
    Single-cluster setups have been produced and analysed.
    In stable configurations, the emergent helium bubbles were considered to be immiscible with the remaining solution. Then, interfacial tension values were determined from pressure tensor inhomogeneities near dividing interfaces. Planar interfaces have been studied as asymptotic limiting cases of zero curvature ($\mathcal{R} \rightarrow \infty$).
   
    At the eutectic point of the lead--lithium system, the interfacial tension at the He/LLA interface, computed by means of the Al-Awad et al. EAM force field~\cite{al2023parametrization}, is consistent with experimental surface tension values of the LLE. This agreement suggests that helium clusters can effectively be treated as bubbles, analogous to those formed at a liquid metal/vacuum interface. 
    In contrast, the EAM model of Belashchenko seems to overestimate interfacial tension values.
    Even so, the two models coherently predict a maxima in $\gamma(x_{\rm Pb})$, both under finite and zero interfacial curvatures.
    
    Helium generally exhibits preferential adsorption on the left side of the interface, which is reflected in local increases in the components of the kinetic pressure tensor. In this region, interatomic interactions are correspondingly stronger. Further, deviations from ideal mixing in lead--lithium alloys are especially pronounced in lithium-rich regions of the phase diagram. These effects are reflected in the maximization of interfacial tension at constant temperature, and are consistent with minimal solubilities at the provided isotherm~\cite{alvarez2025henry}. These effects are also observed in the total $p_N(r)$ profiles, where the expected monotonic (non-increasing) behaviour is locally not satisfied. Curvature of spherical interfaces plays a fundamental role in the determination of the interfacial tension.

    Bubble radii and Gibbs free energy dependence on the solvent composition have also been determined at constant temperature and helium content.
    While the latter resembles interfacial tension profiles, the former is a bit clumsier.
    For BEAM-2019 LM alloys~\cite{belashchenko2019inclusion}, radii show no significant variations compared to AEAM-2023 ones~\cite{al2023parametrization}.
    With the last model, bubbles expand when $x_{\rm Li}/x_{\rm Pb} \simeq 0.19$.
    This behaviour could be related to the so-discussed peculiarity related to the solvent packing~\cite{alvarez2025henry}.

\section*{Acknowledgements}
  
    The authors thank A.Al-Awad for fruitful discussions. J.M. and L.B. acknowledge financial support from the EUROfusion project (HORIZON-101052200-EUROfusion).  J.M.  and E.A.G.  thank financial support of project {PID2024-157478NB-C32} funded by MCIN/AEI/10.13039/5011000-11033 and ERDF `A way of making Europe' by the `European Union NextGenerationEU/PRTR'.
    J.M. gratefully acknowledges financial support from the Generalitat de Catalunya (Grant 2021 SGR 01411).  
    E.A.G. was awarded a fellowship at the Polytechnic University of Catalonia (UPC) by the Spanish `Consejo de Seguridad Nuclear' through the {\it ARGOS} Chair of Nuclear Safety and Radiation Protection.
    E.A.G. also thanks the UPC for financial support through the doctoral grant extension for the fourth year. 

\bibliographystyle{unsrtnat}
\bibliography{references}

\clearpage
\onecolumngrid
\appendix

\section*{Supplementary Information}
\section{Mechanical calculations of pressure tensors}\label{app:A}
    \subsubsection{Stress and pressure tensors}

    The addition of kinetic and virial terms can be expressed as:
    \begin{eqnarray}
        \left\langle \Pi_{\mu\nu} \right \rangle &=& 
        \left\langle \frac{1}{V} \sum_{\alpha, i} m_\alpha (\mathbf{v}_i^\alpha)_\mu (\mathbf{v}_i^\alpha)_\nu  \right\rangle
        \nonumber\\
        && + \left\langle \frac{1}{2V} \sum_{\alpha, i} \sum_{\beta, j} (\mathbf{r}_{ij}^{\alpha\beta})_{\mu} (\mathbf{f}_{ij}^{\alpha\beta})_{\nu} \right\rangle
        \ .
    \end{eqnarray}

    \subsection{Local stress tensors}
        The mechanical approach for interfacial tension calculations is based on the evaluation of local variations of the local stress tensor components, along a series of paths normal to the interface that divides two immiscible phases.

        While the determination of local kinetic terms is somehow trivial, configurational or virial terms are computed by means of the Irving--Kirkwood contour (IKC)~\cite{irving1950statistical}.
        The {IKC} defines the straight path \[\mathbf{r}(\lambda) = (1-\lambda)\mathbf{r}_i^\alpha + \lambda\mathbf{r}_j^\beta = \frac{1}{2}(\mathbf{r}_i^\alpha+\mathbf{r}_j^\beta) + \frac{\lambda'}{2} (\mathbf{r}_j^\beta-\mathbf{r}_i^\alpha)\] joining the positions of all atom pairs, given a free parameter $0 \leq \lambda \leq 1 \Leftrightarrow -1 \leq \lambda' \leq 1$.
        This contour weights the contribution of each particle pair to the virial tensor based on the interface geometry and the orientation of the interparticle forces relative to that interface.
        \begin{itemize}
            \item Considering the first parametrization ($0 \leq \lambda \leq 1$), the {IKC} for planar symmetries can be expressed, accordingly to Refs.~\citenum{ikeshoji2003molecular} and \citenum{lu2022atomistic}, as 
            \begin{eqnarray}
                \left. \frac{\partial\Omega_{ij}^{\alpha\beta}}{\partial z} \right|_p
                &=&
                \lim_{\Delta z \rightarrow 0} \frac{\Delta \Omega_p(z,\Delta z, z_i^\alpha, z_j^\beta)}{\Delta z} 
                \nonumber\\
                &=&
                \frac{1}{|z_{ij}^{\alpha\beta}|} \theta\left(\frac{z-z_i^\alpha}{z_{ij}}\right) \theta\left(\frac{z_j^\beta-z}{z_{ij}}\right)
                .
                \label{eq:IKC_planar}
            \end{eqnarray}
            \item Considering the second parametrization ($-1 \leq \lambda \leq 1$), the {IKC} for spherical symmetries can be expressed, consistently with Ref.~\cite{thompson1984molecular}, as 
            \begin{eqnarray}
                \left. \frac{\partial \Omega_{ij}^{\alpha\beta}}{\partial r} \right|_s &=& \lim_{\Delta r \rightarrow 0} \frac{\Delta \Omega_{ij}^{\alpha\beta}(r,\Delta r, r_i^\alpha, r_j^\beta)}{\Delta r} 
                \nonumber\\
                &=& \frac{1}{r} \Big[\theta(|\lambda_+|-1) +\theta(|\lambda_-|-1) 
                \nonumber\\
                && \qquad - \theta(|\lambda_+|-1)\theta(|\lambda_+|-1)\nonumber\\
                &&\qquad +2\theta(1-|\lambda_+|)\theta(1-|\lambda_-|)\Big] \ ,
                \nonumber
            \end{eqnarray}
            where
            \hspace{5cm}\begin{eqnarray}
                &&\lambda_\pm \equiv \lambda_{min} 
                \nonumber\\
                && \pm \left[{\lambda_{min}}^2 + 1 - 2\left(\frac{(r_i^\alpha)^2+(r_j^\beta)^2}{(r_{ij}^{\alpha\beta})^2}\right) + \frac{4r^2}{(r_{ij}^{\alpha\beta})^2} \right]^{1/2}
                \nonumber
                \ ,
            \end{eqnarray}
            and
            \[
                \lambda_{min} \equiv \dfrac{({r_i^\alpha})^2-({r_j^\beta})^2}{(r_{ij}^{\alpha\beta})^2} \ .
            \]
            The two possible  values of $\lambda_\pm$ determine each of the two possible intersections between an annulus of radius $r$ and the segment that joins $\mathbf{r}_i^\alpha$ to $\mathbf{r}_j^\beta$.
        \end{itemize}

        \subsubsection{Planar symmetries}
        The direction perpendicular to a flat interface is chosen to be the Cartesian coordinate $\mathbf{e}_z$.
        Then, stress tensors are discretized along $z$.
        The interfacial tension calculation method relies on the consideration of a hypothetical strip in the $(y,z)$ plane, of length $L_z$ in the $z$ dimension (and unit length in the $y$ dimension), proposed by Kirkwood and Buff \cite{kirkwood1949statistical}:
        {IT} is defined as the additional stress transferred to the strip beyond the uniform normal stress that would exist in the absence of an interface.

        Pressure tensors and the {HEC} under planar symmetry respectively read~\cite{walton1983pressure}:
        \begin{eqnarray}
            \mathbf{\Pi}(z) = \left[ \mathbf{e}_x \otimes \mathbf{e}_x + \mathbf{e}_y \otimes \mathbf{e}_y \right] p_T (z) \nonumber\\+ \left[ \mathbf{e}_z \otimes \mathbf{e}_z \right] p_N(z) 
            \ \ \ \ \
            \label{eq:stress_planar}
        \end{eqnarray}
        and
        \begin{eqnarray}
            \label{PT_z}
            \frac{\mathrm{d}p_N(z)}{\mathrm{d}z} = 0 
            \label{eq:hydro_planar}
        \end{eqnarray}
        The {HEC} implies that the normal component of the stress tensor, $p_N(z) = p_N(\infty) = \Pi_{zz}$ is constant.
        In other words, the excess pressure in the interface region, $\Delta p(z) = p_N(z) - p_T(z) >0$, must be driven by a decrease in the tangential component with respect to the bulk value ($p_T(z)-p_T(\infty) < 0$), where \[p_T(z) = \frac{1}{2}\left( \Pi_{xx}(z) + \Pi_{yy}(z) \right).\]

        In the vicinities of an interface, densities --and, consequently, kinetic terms of the pressure tensor-- fluctuate along its perpendicular (normal) direction. 
        The density profiles are characterized by particle number densities on differential volumes.
        The density in a planar slice whose normal direction is parallel to the $z$-axis is $\rho(z) = \dfrac{\mathrm{d}n(z)}{\mathrm{d}V}$, where $\mathrm{d}n(z)$ denotes the number of particles that are found within a local slice centred at $z$ of volume $\mathrm{d}V = L_x \ L_y \ \mathrm{d}z$.
        $L_x\equiv x_f - x_i$ and $L_y \equiv y_f - y_i$ denote the lengths of a rectangular-prism simulation box in the $x$ and $y$ axis (respectively), and $\mathrm{d}z$ the differential width of the slice.
        $\theta(x)$ denotes the Heaviside function.
        In the finite-size discretized formulation, the $z$-axis is divided into $M \in \mathbb{Z}$ bins, so that $L_{z}' \equiv z_f - z_i = M \Delta z = N_{int} L_z$.

        \paragraph{Kinetic contributions}
        Kinetic components of the local pressure tensor read
        \[
            [\mathbf{P}_K (z)]_{\mu\nu} \simeq \dfrac{1}{\beta} \left\langle \dfrac{1}{L_x L_y}\lim_{\Delta z \rightarrow0} \dfrac{\Delta n (z)}{\Delta z} \right\rangle \delta_{\mu\nu} \ ,
        \]
        where
        \[
            \dfrac{\Delta n (z)}{\Delta z} = \frac{1}{\Delta z} \sum_{\alpha, i=1}^N \theta\left(z_i^\alpha-z+\frac{\Delta z}{2}\right) \theta\left(z-z_i^\alpha+\frac{\Delta z}{2}\right) \ .
        \]
        \paragraph{Configurational contributions}
        The {IKC} leads to a local pressure tensor for a planar interface in the $x,y$-plane \cite{walton1983pressure}.
        By means of the {IKC}, the fine-grained (discretized along the direction with symmetry, $z$) can be expressed as
        \begin{eqnarray}
            [\mathbf{P}_V(z)]_{\mu\nu}
            = 
            \left\langle \frac{1}{2 L_x L_y} \sum_{\alpha, i=1}^{N_\alpha} \sum_{\beta, j=1}^{N_\beta} [\mathbf{r}_{ij}^{\alpha\beta}]_\mu [\mathbf{f}_{ij}^{\alpha\beta}]_\nu 
            \right. \nonumber\\
            \left.\times\frac{\partial\Omega_p(z,\Delta z, z_i,z_j)}{\partial z} \ \delta_{\mu\nu} \right\rangle
            \\
            \simeq \frac{1}{L_x L_y \Delta z}\left\langle {\sum_{(i,j)}}^* [\mathbf{r}_{ij}]_\mu [\mathbf{f}_{ij}]_\nu \delta_{\mu\nu} \right\rangle
            \ ,
        \end{eqnarray}
        where the factor $1/2$ removes double counting of pairs $(i,j)$.
        {The planar-case {IKC} expression (see equation~\ref{eq:IKC_planar}) represents a coarse-grained Irving--Kirkwood parametrization~\cite{ikeshoji2003molecular}:
        \begin{eqnarray}
            \Delta \Omega(z,{\Delta z}, z_i,z_j) \equiv \frac{z_b(z,z_j,{\Delta z}) - z_a(z,z_i,{\Delta z})}{ z_{ij} } \nonumber\\
            = \stackrel{\text{(see Tab.~\ref{tab:ICK_planar_cases})}}{\dots}
            = \omega(z,{\Delta z},z_i,z_j) \frac{\Delta z}{|z_{ij}|}
        \end{eqnarray}
        where \cite{ikeshoji2003molecular, lu2022atomistic}
        \begin{eqnarray}
            z_c(z,z_k,{\Delta z}) = \left\{\begin{array}{cc}
                z_k , & |z_k-z| < \frac{\Delta z}{2} \\
                z + \frac{\Delta z}{2},  & z_k-z > \frac{\Delta z}{2} \\
                z - \frac{\Delta z}{2},  & z_k-z < -\frac{\Delta z}{2} \\
            \end{array}\right.
        \end{eqnarray}
        and
        \begin{eqnarray}
            \omega(z,{\Delta z},z_i,z_j) &=& \omega_0({\Delta z},z_i,z_j) \  \theta\left(\frac{z-z_i}{z_{ij}}+\frac{\Delta z}{2 |z_{ij}|} \right) \nonumber
            \\
            && \times \theta\left(\frac{z_j-z}{z_{ij}} + \frac{\Delta z}{2 |z_{ij}|}\right)
        \end{eqnarray}
        (being $0 \leq \omega_0(z,z_i,z_j) \leq 1$) calculates the degree of penetration of the pair $i,j$, that cross the finite-size layer $\Delta z$ at $z$, to the virial.
        The IKC defines the fraction of the pair $i,j$ to the total virial.
        \\
        In the limiting case of a layer of infinitesimal width:
        \[\begin{cases}
            \displaystyle\lim_{\Delta z \rightarrow 0} {\Delta \Omega(z,\Delta z,z_i,z_j)} = 
            \lim_{\Delta z \rightarrow 0} \frac{\Delta z}{|z_{ij}|} \theta\left(\frac{z-z_i}{z_{ij}}\right) \theta\left(\frac{z_j-z}{z_{ij}}\right) 
            \\
            \displaystyle\frac{\partial \Omega}{\partial z} = \lim_{\Delta z \rightarrow 0} \frac{\Delta \Omega}{\Delta z} = 
            \frac{1}{|z_{ij}|} \theta\left(\frac{z-z_i}{z_{ij}}\right) \theta\left(\frac{z_j-z}{z_{ij}}\right)
        \end{cases}
        \]
        The product of theta functions turns 1 when the $z$ coordinate is found in between $z_i$ and $z_j$, and is 0 otherwise.
    
        The total pressure tensor equals the mean value of the stress tensor evaluated at all $z$ values:
        \begin{eqnarray}
            P_{\mu\nu} = \bar{\Pi}_{\mu\nu} = \frac{1}{L_z} \int_{-\infty}^{+\infty} \Pi_{\mu\nu}(z) \ \mathrm{d}z \ .
        \end{eqnarray}
        All 9 components of the stress tensor can be summarized in two values: the normal and tangential terms,
        \begin{eqnarray}
            \boxed{\begin{array}{ccc}
                p_T(z) & = & \frac{1}{2}\left[ \Pi_{xx}(z)+\Pi_{yy}(z) \right] \ ,  \\
                p_N(z) & = & p_N(\infty) = \Pi_{zz} = constant 
            \end{array}}
        \end{eqnarray}
        respectively.
        They must satisfy:
        \begin{itemize}
            \item \textit{Homogeneity conditions}
            We assume that the simulation box extends infinitely along the $x,y$ directions (in practice, it is implemented using PBC). That means that all $x,y$ points become equivalent after ensemble averaging, i.e., there is homogeneity in both $x$ and $y$ axis.
            Therefore $\forall (x,y,z)\in\mathbb{R}^3$ and $\mu,\nu\in\{x,y,z\}$ (finite volume approximation):
            \begin{eqnarray}
                \left.\begin{array}{c}
                    \displaystyle \lim_{{\Delta x}\rightarrow0} \Pi_{\mu\nu}(x+\frac{\Delta x}{2},y,z) = \lim_{{\Delta x}\rightarrow0} \Pi_{\mu\nu}(x-\frac{\Delta x}{2},y,z) \nonumber
                    \\
                    \displaystyle \Rightarrow \frac{\partial \Pi_{\mu\nu}}{\partial x} = 0
                \end{array}\right\}
                \\
                \left.\begin{array}{c}
                    \displaystyle \lim_{{\Delta y}\rightarrow0} \Pi_{\mu\nu}(x,y+\frac{\Delta y}{2},z) = \lim_{{\Delta y}\rightarrow0} \Pi_{\mu\nu}(x,y-\frac{\Delta y}{2},z) \nonumber
                    \\
                    \displaystyle \Rightarrow \frac{\partial \Pi_{\mu\nu}}{\partial y} = 0
                \end{array}\right\}
                \\
                \left.\begin{array}{c}
                    \displaystyle \lim_{{\Delta z}\rightarrow0} \Pi_{\mu\nu}(x,y,z+\frac{\Delta z}{2}) \stackrel{(?)}{=} \lim_{{\Delta z}\rightarrow0} \Pi_{\mu\nu}(x,y,z-\frac{\Delta z}{2}) \nonumber
                    \\
                    \displaystyle \Rightarrow \frac{\partial \Pi_{\mu\nu}}{\partial z} \stackrel{(?)}{=} 0
                \end{array}\right\}
            \end{eqnarray}
            All components of the stress tensor can depend only on $z$, $\Pi_{\mu\nu} = \Pi_{\mu\nu}(z)$.
    
            \item \textit{Net force balance on diagonal components} 
            At the same time, assuming hydrostatic balance (there is no net flow), the diagonal components of the stress tensor on one face of the infinitesimal cube must balance by the component of the opposite face.
            Net force balance for $\Pi_{xx}(z)$ and $\Pi_{yy}(z)$ is trivially ensured by the homogeneity conditions, irrespective of the value of $z$.
            The HEC imposes an additional restriction:
            \begin{eqnarray}
                \Pi_{zz}(z-\frac{\Delta z}{2}) = \Pi_{zz}(z+\frac{\Delta z}{2})
            \end{eqnarray}
            Therefore, $\Pi_{zz}$ depends neither on $x$, $y$ nor $z$.
            The component of the stress tensor in the direction normal to the interface, then, equals to the same at bulk conditions (very far from the interface).
    
            \item \textit{Isotropy in the $x-y$ plane}
            $\Pi_{xx}(z) = \Pi_{yy}(z)$.
    
            \item \textit{Net force balance on off-diagonal components}
            Off-diagonal components, however, $\Pi_{\mu,\nu\neq\mu}(z)$, must be zero, since there is no counterforce that could balance a non-zero contribution.
        \end{itemize}
            
        \begin{table*}[h!] 
            \centering
            \caption{Summary of {IKC} cases for planar symmetries. 
            }
            \label{tab:ICK_planar_cases}
            \rotatebox{90}{
            \begin{tabular}{|c|c|c|c|c|c|c|c|c|}
                \hline
                \multicolumn{2}{|c|}{CASE} & CONDITION & $z_a$ & $z_b$ & $z_{ij} \equiv z_j-z_i$ & \multicolumn{2}{|c|}{$\Delta \Omega(z,z_i,z_j)$} & $\omega({z,\Delta z},z_i,z_j)$\\\hline
                \multirow{2}{*}{1} & (1A) & $z_i+\Delta z/2 < z < z_j-\Delta z/2$ & $z-\Delta z/2$ & $z+\Delta z/2$ & $>0$ & $\dfrac{\Delta z}{z_{ij}}$ & \multirow{2}{*}{$ \dfrac{\Delta z}{|z_{ij}|}$} & \multirow{2}{*}{$1$} \\
                ~ & (1B) & $z_j+\Delta z/2 < z < z_i-\Delta z/2$ & $z+\Delta z/2$ & $z-\Delta z/2$ & $<0$ & $-\dfrac{\Delta z}{z_{ij}}$ & ~ & ~ \\\hline
                \multirow{2}{*}{2} & (2A) & $z-\Delta z/2 < z_i < z_j < z + \Delta z/2$ & $z_i$ & $z_j$ & $>0$ & $\dfrac{z_j-z_i}{|z_{ij}|}$
                & \multirow{2}{*}{$ 1$} & \multirow{2}{*}{$\dfrac{z_{ij}}{\Delta z}$}\\
                ~ & (2B) & $z-\Delta z/2 < z_j < z_i < z + \Delta z/2$ & $z_i$ & $z_j$ & $<0$ & $-\dfrac{z_j-z_i}{|z_{ij}|}$ & ~ & ~ \\\hline
                \multirow{2}{*}{3} & (3A)  & $z-\Delta z/2 < z_i < z+\Delta z/2<z_j$ & $z_i$ & $z+\Delta z/2$ & $>0$ & $\dfrac{z+\Delta z/2 - z_i}{|z_{ij}|}$  
                & \multirow{2}{*}{\dots} & $+\dfrac{z+\Delta z/2 - z_i}{\Delta z}$\\
                ~ & (3B) & $z_j<z-\Delta z/2 < z_i < z+\Delta z/2$ & $z_i$ & $z-\Delta z/2$ & $<0$  & $-\dfrac{z-\Delta z/2 - z_i}{|z_{ij}|}$ 
                & ~ & $-\dfrac{z-\Delta z/2 - z_i}{\Delta z}$ \\\hline
                \multirow{2}{*}{4} & (4A) & $z-\Delta z/2 < z_j < z+\Delta z/2<z_i$  & $z+\Delta z/2$ & $z_j$ & $<0$ & $-\dfrac{z_j-z-\Delta z/2}{|z_{ij}|}$ 
                & \multirow{2}{*}{\dots} & $-\dfrac{z_j-z-\Delta z/2}{\Delta z}$ \\
                ~ & (4B) & $z_i<z-\Delta z/2 < z_j < z+\Delta z/2$ & $z-\Delta z/2$ & $z_j$ & $>0$  & $\dfrac{z_j-z+\Delta z/2}{|z_{ij}|}$ 
                & ~ & $\dfrac{z_j-z+\Delta z/2}{\Delta z}$ \\\hline
                \multirow{4}{*}{5} & (5A) & $z_i < z_j < z-\Delta z/2 $
                & $z-\Delta z/2$ & $z-\Delta z/2$ & $>0$ 
                & \multirow{4}{*}{$0$} 
                & \multirow{4}{*}{$0$} & \multirow{4}{*}{$0$}\\
                ~ & (5B) & $z_j<z_i<z-\Delta z/2$ & $z-\Delta z/2$ & $z-\Delta z/2$ & $<0$ & ~ & ~ & ~ \\
                ~ & (5C) & $z+\Delta z/2 < z_i < z_j$ & $z+\Delta z/2$ & $z+\Delta z/2$ & $>0$ & ~ & ~ & ~ \\
                ~ & (5D) & $z+\Delta z/2<z_j<z_i$ & $z+\Delta z/2$ & $z+\Delta z/2$ & $<0$ & ~ & ~ & ~ \\
                \hline
                \multicolumn{9}{c}{~}
                \\
                \multicolumn{9}{l}{(CASE 3) $\Leftrightarrow|z_i-z|<\Delta z / 2 \text{ and } |z_j-z|>\Delta z/2; \qquad$}
                \\
                \multicolumn{9}{l}{(CASE 4) $\Leftrightarrow|z_i-z|>\Delta z / 2 \text{ and } |z_j-z|<\Delta z/2$}
            \end{tabular}}
        \end{table*}
        }
        \subsubsection{Spherical symmetries}
        For spherically symmetric interfaces, stress tensors are discretized along radial directions, $\mathbf{e}_r$.
        In addition, isotropy is assumed, so contributions are averaged over all possible angular orientations.
        Then, stress tensor components become functions strictly of the radial coordinate, $r \equiv |\mathbf{r}-\mathbf{r}_{CoM}|^*$, with origin at the centre of masses of the spherical phase. The superindex $*$ indicates that distances take into account {periodic boundary conditions (PBC)}.

        Similarly, stress tensors and the {HEC} under spherical symmetry read~\cite{thompson1984molecular}:
        \begin{eqnarray}
            &&\mathbf{\Pi}(r) = \left[ \mathbf{e}_\phi \otimes \mathbf{e}_\phi + \mathbf{e}_\theta \otimes \mathbf{e}_\theta \right] p_T (r) + \left[ \mathbf{e}_r \otimes \mathbf{e}_r \right] p_N(r)
            \ \ \ \ \
            \label{eq:stress_spherical}
        \end{eqnarray}
        and
        \begin{eqnarray}
            &&\frac{\mathrm{d}p_N(r)}{\mathrm{d}r} + \frac{2}{r} {\Delta p}_s(r) = 0  \ ,
            \label{eq:hydro_spherical}
        \end{eqnarray}
        respectively; 
        where we made use of the definition
        $
        {\Delta p}_s(r) \equiv p_N(r) - p_T(r)
        $.
        The interfacial tension at the interface of tension is assumed to satisfy the Young--Laplace equation \cite{thompson1984molecular, montero2020interfacial}.
        
        \paragraph{Kinetic contribution}
            Kinetic terms of local pressure tensors account for the products of velocity components of all particles that lie into a finite region of volume $\mathrm{d}V = A(r) \ \mathrm{d}r$ located in the centre of a shell at $r$,
            \begin{eqnarray}
                K_{\mu\nu}(r) 
                = \left\langle \sum_{\alpha, i} m_\alpha (\mathbf{v}_i^\alpha)_\mu (\mathbf{v}_i^\alpha)_\nu  \ \Theta \left(r,r_i^\alpha+\frac{\mathrm{d}r}{2},\ r_i^\alpha-\frac{\mathrm{d}r}{2}\right)\right\rangle 
            \end{eqnarray}
            where $r_i^\alpha \equiv |\mathbf{r}_i^\alpha - \mathrm{r}_{CoM}|^*$, and 
            \[
                \Theta (x,a,b) \equiv \theta\left(a-x\right) \ \theta\left(x-b\right)
            \]
            denotes the product of two Heaviside functions with opposite sign,
            $A(r) = 4\pi r^2$ is the area of a shell of width $\mathrm{d}r$., and $(r-\mathrm{d}r/2,r+\mathrm{d}r/2)$ are its inner and outer boundaries.

            As in {molecular dynamics} simulations the velocities are determined by the temperature of the thermostat, the latter average can be simply approximated to $K_{\mu \nu} \simeq \beta^{-1} n(r) \delta_{\mu\nu}$, where $n(r)$ is the number of particles within the shell. 
            In other words, the pressure tensor is diagonal:
            \begin{equation}
                \mathbf{P}_K(r)  \simeq \beta^{-1} \langle \rho(r) \rangle \  \mathbf{I} \ ,
                \label{eq:PKr}
            \end{equation}
            where $\mathbf{I}$ denotes the identity tensor,
            and
            \[
                \rho(r) = \frac{\mathrm{d}n(r)}{\mathrm{d}V} = \frac{1}{4\pi r^2} \frac{\mathrm{d}n(r)}{\mathrm{d}r} = \sum_\alpha \rho_\alpha(r)
            \]
            denotes radial density fluctuations with respect to the centre of the spherical phase.
            From a computational perspective, instantaneous radial density profiles are determined as the ratio between: (a) the number of atoms enclosed in a shell of thickness $\Delta r$, whose midpoint is found at $r$ ; and (b) the volume of the shell~\cite{thompson1984molecular}:
            \begin{eqnarray}
                \rho (r) \simeq \frac{\Delta n(r)}{4\pi (r^2+\Delta r/12) \Delta r} \ .
            \end{eqnarray}

    \paragraph{Virial contribution}
    The normal component of the configurational (virial) contribution to radial pressure tensors reads:
    \begin{eqnarray}
        && [\mathbf{P}_V (r)]_{N} = \mathbf{e}_r \cdot \mathbf{P}_V(r) \cdot \mathbf{e}_r 
        \nonumber\\
        && = \left\langle \frac{1}{4 \pi r^2} \frac{1}{2}\sum_{\alpha, i=1}^{N_\alpha} \sum_{\beta, j=1}^{N_\beta} 
        \left| \mathbf{r} \cdot \mathbf{f}_{ij}^{\alpha\beta} \right| \left.\ \frac{\partial\Omega_{ij}^{\alpha\beta}}{\partial r}\right|_s
        \right\rangle
        \ ,
        \label{eq:PVr}
    \end{eqnarray}
    where $r = |\mathbf{r}|$
    is the distance to the centre of mass of the sphere;  $\left| \frac{\mathbf{r}}{r} \cdot \mathbf{f}_{ij}^{\alpha\beta} \right| = \left| \frac{\mathbf{r}}{r} \cdot f(r_{ij}^{\alpha\beta}) \frac{\mathbf{r}_{ij}^{\alpha\beta}}{r_{ij}^{\alpha\beta}} \right| = |[\mathbf{f}_{ij}]_r^{\alpha\beta}|$ is the projection of the pair force $\mathbf{f}_{ij}$ to the normal direction $\hat{\mathbf{r}} \equiv \mathbf{r}/r$; 
    \begin{eqnarray}
        \left.\ \frac{\partial\Omega_{ij}^{\alpha\beta}}{\partial r}\right|_s
        &=& \lim_{\Delta r \rightarrow 0} \frac{\Delta \Omega_s(r,{\Delta r},r_i,r_j)}{\Delta r}
        \nonumber\\
        &=& \frac{1}{r} 
        \left[ \theta(|\lambda_+|-1) + \theta(|\lambda_-|-1) \right.
        \nonumber\\
        && \ \ \ \  - \left. \theta(|\lambda_+|-1)\times\theta(|\lambda_-|-1) \right. \nonumber\\
        && \ \ \ \  \left.+ 2\times\theta(1-|\lambda_+|)\times\theta(1-|\lambda_-|) \right]
        \ . \ \ 
        \label{eq:numintersecs}
    \end{eqnarray}
    represents the {IKC} 
    and $\displaystyle {\sum_{\stackrel{\alpha, \beta}{(i,j)}}}'$ denotes the restricted sum over pair forces that cross a surface of radius $r$.
    All five possible scenarios have been represented in Tables~\ref{rep:1-2}--\ref{rep:4-5}.

        \paragraph{Total local pressure tensor}
        Under spherical symmetry, the discrete pressure tensor normal and tangential components respectively read:
        \begin{eqnarray}
            p_N(r)  &=&  P_{rr}(r) = \mathbf{e}_r \cdot \mathbf{P}(r) \cdot \mathbf{e}_r 
            \nonumber\\
            &=& 
            \lim_{\Delta r\rightarrow0} \frac{1}{4\pi r^2 \ {\Delta r} }\Bigg[ \langle \Delta n(r) \beta^{-1} \rangle 
            \nonumber\\
            && + \frac{1}{2}\Bigg\langle \sum_{\alpha, i} \sum_{\beta, j} [\mathbf{r}_{ij}^{\alpha\beta}]_\mu [\mathbf{f}_{ij}^{\alpha\beta}]_\nu \ {\Delta \Omega}_{ij}^{\alpha\beta} \Bigg\rangle \Bigg]
            \nonumber\\
            &\simeq& 
            \Bigg[\left\langle\rho(r) \beta^{-1}\right\rangle 
            \nonumber\\
            &&+ \frac{1}{2}\Bigg\langle \sum_{\alpha, i} \sum_{\beta, j} [\mathbf{r}_{ij}^{\alpha\beta}]_\mu [\mathbf{f}_{ij}^{\alpha\beta}]_\nu \  \frac{\partial \Omega_{ij}^{\alpha\beta}}{\partial r} \Bigg\rangle \Bigg]
            \ .
            \label{eq:PNr}
        \end{eqnarray}
        and
        \begin{eqnarray}
            p_T(r) &=& \frac{1}{2} \left[P_{\theta\theta}(r) + P_{\phi\phi}(r)\right] \nonumber\\
            &=& \frac{1}{2} \left[\mathbf{e}_\theta \cdot \mathbf{P}(r) \cdot \mathbf{e}_\theta \ + \ \mathbf{e}_\phi \cdot \mathbf{P}(r) \cdot \mathbf{e}_\phi\right]
            \nonumber\\
            &=& 
            p_N(r) + \frac{r}{2} \frac{\mathrm{d}p_N(r)}{\mathrm{d}r}
            \ .
            \label{eq:PTr}
        \end{eqnarray}
        While Eq.~\ref{eq:PNr} straightly accounts for the sum of isotropic mechanical contributions (\ref{eq:PKr} and \ref{eq:PVr}),
        Eq.~\ref{eq:PTr} is a direct consequence of hydrostatic equilibrium (\ref{eq:hydro_spherical}).

    \clearpage
\begin{table*}[]
    \centering
    \caption{Representation of the IKC for a given pair of particles ($i,j$) and a spherical shell of radius $r$ -- Cases 1 and 2.}
    \label{rep:1-2}
    \begin{tabular}{ccc}
        Case & Schematic Representation 
        \\ \hline\hline
        (1) \\
        \begin{minipage}{15em}
        \begin{itemize}
            \item $i,j$ outside the circle;
            \item The straight line that joins $i$ and $j$ does not intersect with the surface of radius $r$;
            \item No intersections ($\lambda \notin \mathbb{R}$)
        \end{itemize}
        \end{minipage} &
        \begin{minipage}{20em}
            \begin{tikzpicture}
              \coordinate (O) at (0,0);
              \draw (O) circle (2.5cm);
              \draw (O) [line width=.5pt,dash pattern=on 1pt off 2pt,color=green]circle(2.6cm);
              \draw (O) [line width=.5pt,dash pattern=on 1pt off 2pt, color=orange]circle (2.4cm);
              \node at (-30:3.3) {\color{green}$r+\Delta r /2$};
              \node at (-60:1.8) {\color{orange}$r-\Delta r /2$};
            
              \coordinate (i) at (128.85:4);
              \filldraw [] (i) circle (0.05cm);
              \coordinate (j) at (34.15:4);
              \filldraw [] (j) circle (0.05cm);
              \coordinate (i1) at (142.85:4);
              \filldraw [] (i1) circle (0.05cm);
              \coordinate (j1) at (136.15:3);
              \filldraw [] (j1) circle (0.05cm);
              \coordinate (i2) at (27.85:3);
              \filldraw [] (i2) circle (0.05cm);
              \coordinate (j2) at (22.15:4);
              \filldraw [] (j2) circle (0.05cm);
              
              \draw[thick, ->, dotted] (i) -- (j);
              \draw[thick, ->, dotted] (i1) -- (j1);
              \draw[thick, ->, dotted] (i2) -- (j2);
            
              \node[left] at (i1) {i'};
              \node[right] at (j1) {j'};
              \node[left] at (i2) {i''};
              \node[right] at (j2) {j''};
              \node[left] at (i) {i};
              \node[right] at (j) {j};
        \end{tikzpicture}
        \end{minipage}
        \\
        ~
        \\\hline
        (2)\\
        \begin{minipage}{15em}
        \begin{itemize}
            \item One particle inside and the other outside the circle.
            \item One intersection (only A contributes):
            \[\begin{cases}
                |\lambda_A| \leq 1
                \\
                |\lambda_B| > 1
            \end{cases}\]
        \end{itemize}
        \end{minipage}&
        \begin{minipage}{20em}
        \begin{tikzpicture}
          \coordinate (O) at (0,0);
          \draw (O) circle (2.5cm);
          \draw (O) [line width=.5pt,dash pattern=on 1pt off 2pt,color=green]circle(2.6cm);
          \draw (O) [line width=.5pt,dash pattern=on 1pt off 2pt, color=orange]circle (2.4cm);
        
          \coordinate (A) at (120:2.5);
          \coordinate (B) at (45:2.5);
        
          \coordinate (i) at (142.85:4);
          \filldraw [] (i) circle (0.05cm);
          \coordinate (j) at (106.210445:2.16123300818374);
          \filldraw [] (j) circle (0.05cm);
          \coordinate (i1) at (58.7895549678057:2.16123300818374);
          \coordinate (j1) at (22.15:4);
          \draw[->, thick] (O) -- (A) node[midway, left] {\small$\mathbf{r}_A$};
          \draw[->, thick] (O) -- (B) node[midway, right] {\small$\mathbf{r}_B$};
        
          \draw[thick, -, dotted] (i) -- (j1);
        
          \draw[->, thick] (i) -- (j) node[midway, below, yshift=-.1cm] {\small$\displaystyle {\mathbf{r}_{ij}}$};
        
          \node[left] at (i) {i};
          \node[right] at (j) {j};
          \node at (120:2.8) {\color{red}A};
          \node at (45:2.8) {\color{blue}B};
          \node at (-90:3.5) {$r = |\mathbf{r}_A| = |\mathbf{r}_B|$};
          \filldraw [red] (120:2.5) circle (.05cm);
          \filldraw [blue] (45:2.5) circle (.05cm);

          \node at (-30:3.3) {\color{green}$r+\Delta r /2$};
          \node at (-60:1.8) {\color{orange}$r-\Delta r /2$};
        \end{tikzpicture}
        \end{minipage}
        \\
        ~
        \\\hline\hline
    \end{tabular}
\end{table*}

\begin{table*}[]
    \centering
    \caption{Representation of the IKC for a given pair of particles ($i,j$) and a spherical shell of radius $r$ -- Case 3 and 4.}
    \label{rep:4-5}
    \begin{tabular}{ccc}
        Case & Schematic Representation 
        \\ \hline\hline
        (4)\\
        \begin{minipage}{15em}
        \begin{itemize}
            \item Both $i,j$ inside the circle.
            \item 2 intersections at $\mathbf{r}_A, \mathbf{r}_B$ ($\lambda$ has two real solutions) that DO NOT contribute.
        \end{itemize}
        \end{minipage}
        & 
        \begin{minipage}{20em}                 
        \begin{tikzpicture}
              \coordinate (O) at (0,0);
              \draw (O) circle (2.5cm);
              \draw (O) [line width=.5pt,dash pattern=on 1pt off 2pt,color=green]circle(2.6cm);
              \draw (O) [line width=.5pt,dash pattern=on 1pt off 2pt, color=orange]circle (2.4cm);
            
              \coordinate (A) at (120:2.5);
              \coordinate (B) at (45:2.5);
            
              \coordinate (i) at (142.85:4);
              \coordinate (j) at (22.15:4);
              \coordinate (i1) at (106.210445:2.16123300818374);
              \filldraw [] (i1) circle (0.05cm);
              \coordinate (j1) at (58.7895549678057:2.16123300818374);
              \filldraw [] (j1) circle (0.05cm);
              \draw[->, thick] (O) -- (A) node[midway, left] {\small$\mathbf{r}_A$};
              \draw[->, thick] (O) -- (B) node[midway, right] {\small$\mathbf{r}_B$};
            
              \draw[thick, -, dotted] (i) -- (j);
            
              \draw[->, thick] (i1) -- (j1) node[midway, below, yshift=-.1cm] {\small$\displaystyle {\mathbf{r}_{ij}}$};
            
              \node[left] at (i) {i};
              \node[right] at (j) {j};
              \node at (120:2.8) {\color{red}A};
              \node at (45:2.8) {\color{blue}B};
              \node at (-90:3.5) {$r = |\mathbf{r}_A| = |\mathbf{r}_B|$};
              \filldraw [red] (120:2.5) circle (.05cm);
              \filldraw [blue] (45:2.5) circle (.05cm);

              \node at (-30:3.3) {\color{green}$r+\Delta r /2$};
              \node at (-60:1.8) {\color{orange}$r-\Delta r /2$};
            \end{tikzpicture} 
        \end{minipage}
            \\ ~  \\
        \hline
        (5)\\
        \begin{minipage}{15em}
        ~ \\
        \begin{itemize}
            \item Both $i,j$ outside the circle;
            \\ ~ \\
            \item 2 intersections at $\mathbf{r}_A, \mathbf{r}_B$ ($\lambda$ has two real solutions) that contribute equally:
            \\ ~ \\
            \vspace{3cm}
                \[
                    |\mathbf{r}\cdot(\mathbf{r}_j-\mathbf{r}_i)| = \frac{{r_{ij}}^2}{2} \sqrt{\left(\frac{r_i^2-r_j^2}{r_{ij}^2}\right)^2 + 1 - 2 \left(\frac{r_i^2+r_j^2}{r_{ij}^2}\right)+\frac{4r^2}{r_{ij}^2}} \ .
                \]
        \end{itemize}
        \end{minipage}
        & 
        \begin{minipage}{20em}
        \begin{tikzpicture}
              \coordinate (O) at (0,0);
              \node at (90:3.5){\vspace{-2cm}};
              \draw (O) circle (2.5cm);
              \draw (O) [line width=.5pt,dash pattern=on 1pt off 2pt,color=green]circle(2.6cm);
              \draw (O) [line width=.5pt,dash pattern=on 1pt off 2pt, color=orange]circle (2.4cm);
            
              \coordinate (A) at (120:2.5);
              \coordinate (B) at (45:2.5);
            
              \coordinate (i) at (142.85:4);
              \filldraw [] (i) circle (0.05cm);
              \coordinate (j) at (22.15:4);
              \filldraw [] (j) circle (0.05cm);
              \draw[->, thick] (O) -- (A) node[midway, left] {\small$\mathbf{r}_A$};
              \draw[->, thick] (O) -- (B) node[midway, right] {\small$\mathbf{r}_B$};
            
              \draw[thick, ->, dotted] (i) -- (j);
            
              \draw[->, thick] (A) -- (B) node[midway, below, yshift=-.1cm] {\small$\displaystyle {\mathbf{r}_{ij}}$};
            
              \node[left] at (i) {i};
              \node[right] at (j) {j};
              \node at (120:2.8) {\color{red}A};
              \node at (45:2.8) {\color{blue}B};
              \node at (90:3.5) {$r = |\mathbf{r}_A| = |\mathbf{r}_B|$};
              \node at (-90:4) {~};
              \filldraw [red] (120:2.5) circle (.05cm);
              \filldraw [blue] (45:2.5) circle (.05cm);

              \node at (-30:3.3) {\color{green}$r+\Delta r /2$};
              \node at (-60:1.8) {\color{orange}$r-\Delta r /2$};

            \end{tikzpicture}\end{minipage}
            \\
            ~
            \\
            \hline\hline
    \end{tabular}
    \label{tab:my_label}
\end{table*}
\begin{table*}[]
    \centering
    \caption{Representation of the IKC for a given pair of particles ($i,j$) and a spherical shell of radius $r$ -- Case 5.}
    \label{rep:3}
    \begin{tabular}{ccc}
        Case & Schematic Representation 
        \\ \hline\hline
        (3) \\
        \begin{minipage}{15em}
            \begin{itemize}
                \item Both $i,j$ outside the circle
                \item The positions of A and B collapse, and the two intersections collapse into a single one: $\lambda_A = \lambda_B$ (degeneration).
                \item These terms do not contribute as the intersection position vector is perpendicular to the straight line that joins $i$ and $j$: \[|\mathbf{r}_{A,B} \ \cdot \ (\mathbf{r}_{j}-\mathbf{r}_i)| = 0 \ . \]
            \end{itemize}
        \end{minipage}&
        \begin{minipage}{20em}
        \begin{tikzpicture}
          \coordinate (O) at (0,0);
          \draw (O) circle (2.5cm);
          \draw (O) [line width=.5pt,dash pattern=on 1pt off 2pt,color=green]circle(2.6cm);
          \draw (O) [line width=.5pt,dash pattern=on 1pt off 2pt, color=orange]circle (2.4cm);
        
          \coordinate (A) at (84:2.5);
          \coordinate (B) at (84:2.5);
        
          \coordinate (i) at (132.273689006094:3.71651718682963);
          \filldraw [] (i) circle (0.05cm);
          \coordinate (j) at (42.2736890060937:3.4);
          \filldraw [] (j) circle (0.05cm);
          \draw[->, thick] (O) -- (A) node[midway, left] {\small$\mathbf{r}_A$};
          \draw[->, thick] (O) -- (B) node[midway, right] {\small$\mathbf{r}_B$};
        
          \draw[thick, -, dotted] (i) -- (j);
        
          \draw[->, thick] (i) -- (j) node[midway, above, yshift=-.1cm] {\small$\displaystyle {\mathbf{r}_{ij}}$};
        
          \node[left] at (i) {i};
          \node[right] at (j) {j};
          \node at (88:2.3) {\color{red}A};
          \node at (80:2.3) {\color{blue}B};
          \node at (-90:3.5) {$r = |\mathbf{r}_A| = |\mathbf{r}_B|$};
          \filldraw [red] (84:2.5) circle (.05cm);
          \filldraw [blue] (84:2.5) circle (.025cm);

          \node at (-30:3.3) {\color{green}$r+\Delta r /2$};
          \node at (-60:1.8) {\color{orange}$r-\Delta r /2$};
        \end{tikzpicture}
        \end{minipage}
        \\
        ~
        \\\hline\hline
    \end{tabular}
\end{table*}
\clearpage
    
    \clearpage
    \subsection{Interfacial tension}
        The interfacial tension intrinsically arises from a non-zero normal--tangential pressure difference; namely, a finite local stress anisotropy.
        Along transverse directions, pressure components are always assumed to remain independent of the presence of any interface, and their values equal both the total pressure ($p$, imposed by the barostat) and the normal component very far from the interface, $p_N(\infty)$.
        Obviously, when there is no interface, $\gamma = 0$, and there is no normal--tangential pressure difference ($p_N(r) = p_T (r) = p$).

        Consistent with the notation used in Eq.~\ref{eq:stress_sym}, we define $n$-moments for given arbitrary symmetries as 
        \begin{eqnarray}
            \langle {(\xi_N)}^n \rangle_{sym} &=& \int_{D_x} {(\xi_N)}^n f_{sym}(\xi_N) \ \mathrm{d}\xi_N
        \end{eqnarray}
        where 
        $\xi_N$ is the coordinate along which the system shows the symmetry, and
        $f_{sym}(x)$ is a normalized distribution function of the latter, extended over the domain $D_f$, which is related to the pressure difference \[ \Delta p(\xi_N) \equiv p_N(\xi_N) - p_T(\xi_N) \ ,\]
        which measures local anisotropies of the pressure in directions perpendicular to the interface of tension.

        In addition, we define the function
        \begin{eqnarray}\label{eq:Gamma_mom}
            \Gamma_n(\xi_N) &\equiv& \int_{D_x} \left({\frac{\xi_N}{\Lambda_{sym}}}\right)^n {\Delta p}(\xi_N) \ \mathrm{d}\xi_N \ .
        \end{eqnarray}
        Considering $\Lambda_{sym}$ an arbitrary constant factor with units of length, $\Gamma_n$ is measured in units of interfacial tension.
        In following, we will see that:
        \begin{enumerate}
            \item Planar symmetries: 
            \[
            \begin{cases}
                \displaystyle\langle z \rangle_p = \frac{1}{\gamma_p}\int_{-\infty}^\infty \mathrm{d}z \ z\ [p_N(z)-p_T(z)] \\
                \displaystyle\qquad= \frac{1}{L_z' \ (\Delta p)_p}\int_{-\infty}^\infty \mathrm{d}z \ z\ [p_N(z)-p_T(z)] 
                \\
                \displaystyle\qquad= \frac{1}{N_{int}} \sum_{s=1}^{N_{int}} z_s
                \\
                \Lambda_p = L_z' = N_{int} L_z
                \\
                \displaystyle \gamma_p = \Gamma_{n=1}(z) = L_z' \ (\Delta p)_p
            \end{cases}                
            \]
            \item Spherical symmetries:
            \[
            \begin{cases}
                \displaystyle
                (\mathcal{R}_s)^3 = \langle r^3 \rangle_s = -\frac{1}{(\Delta p)_s}\int_0^\infty \mathrm{d}r \ r^3 \ \dfrac{\mathrm{d}p_N}{\mathrm{d}r} \\
                \displaystyle\qquad= \frac{2}{(\Delta p)_s} \int_0^\infty \mathrm{d}r \ r^2 \ [p_N(r) - p_T(r)]
                \\
                \displaystyle\qquad\quad = \dfrac{\mathcal{R}_s}{\gamma_s} \int_0^\infty \mathrm{d}r \ r^2 \ [p_N(r) - p_T(r)]
                \\
                \displaystyle \Lambda_s = \mathcal{R}_s
                \\
                \displaystyle\gamma_s = \Gamma_{n=2}(r)
                = \dfrac{1}{(\mathcal{R}_s)^2} \int_0^\infty \mathrm{d}r \ r^2 \ [p_N(r) - p_T(r)] 
                \\
                \displaystyle \quad = \left[-\dfrac{{(\Delta p)_s}^{2}}{8} \int_0^\infty \mathrm{d}r \ r^3 \ \dfrac{\mathrm{d}p_N(r)}{\mathrm{d}r}\right]^{1/3}
                = \dfrac{(\Delta p)_s \ \mathcal{R}_s}{2}
            \end{cases}                
            \]
        \end{enumerate}

        \subsubsection{Planar interfaces}
        Walton et al. already pointed out that, given a planar division between two surfaces, the \textit{surface tension}
        and the \textit{surface of tension} correspond to the 0-th and 1-st moment of the difference ${\Delta p}_p (z) \equiv p_N(z)-p_T(z) \equiv [p_N-p_T]_p(z)$ \cite{walton1983pressure}.
        Then, the definition of the pseudo-distribution function
        reads:
        \begin{eqnarray}\label{f_p}
            f_p(z) \equiv \frac{1}{\gamma_p} {\Delta p}_p (z)
            \ .
        \end{eqnarray}
        It seems an appropriate choice, consistent with
        \begin{eqnarray}
            \frac{1}{N_{int}} \sum_{s=1}^{N_{int}} z_s = \frac{\int_{-\infty}^\infty z \ \ \mathrm{d}z \ \ {\Delta p}_p(z)}{\int_{-\infty}^\infty \mathrm{d}z \ \ {\Delta p}_p(z)} = \bar{z}  = \langle z \rangle_p
            \ .
        \end{eqnarray}
        The term of `pseudo-distribution' is not accidental, but in Ref.~\citenum{walton1983pressure} (using both Irving--Kirkwood and Harashima contours) one readily realizes that $[p_N-p_T](z)$ is positive-defined for most of the values of $z$, but (small) negative values may be possible in finite regions near the surfaces of tension.
        Aside from that, the definition of the positive overall pressure difference~\cite{yang2014diffusion, lu2022atomistic}
        \begin{eqnarray}
            (\Delta p)_p = 
            \frac{\gamma_p}{L_z'} = \frac{1}{N_{int} L_z} \int_{-\infty}^\infty \mathrm{d}z \ \ {\Delta p}(z)
            \label{eq:dif_press_planar_1}
            \\
            = \bar{\Pi}_{zz} - \frac{1}{2}[\bar{\Pi}_{xx}+\bar{\Pi}_{yy}]> 0
            \label{eq:dif_press_planar_2}
        \end{eqnarray}
        ensures that the distribution is normalized to unity.
        It means that, though the normal-tangential pressure difference is allowed to be negative locally negative, the global difference always must be positive.
        Otherwise, it would lead to (non-physical) negative interfacial tension values.
    
        In Eq.~\ref{eq:dif_press_planar_2}, diagonal global-mean pressure-tensor components consistently read:
        \[
            \bar{\Pi}_{\mu \mu} = \frac{1}{N_{int} L_z} \int_{-\infty}^\infty \mathrm{d}z \ \Pi_{\mu\mu}(z) \ ,
        \]
        which yields to the approximated `virial-method' expression for the planar-case interfacial tension~\cite{lu2022atomistic}:
        \begin{eqnarray}
            \gamma_p \simeq \frac{1}{L_x L_y} \left\langle\bar{S}_{zz} - \frac{1}{2}\left(\bar{S}_{xx}+\bar{S}_{yy}\right)\right\rangle \ .
        \end{eqnarray}

        \subsubsection{Spherical interfaces}
        Analogously to the definition of the $n$-th moment of $f_p(z)$ in the planar-symmetry case, we define --under spherical symmetry-- both the pseudo-distribution
        \begin{eqnarray}\label{f_s}
            f_s(r) &\equiv& -\frac{1}{(\Delta p)_s} \frac{\mathrm{d}p_N(r)}{\mathrm{d}r} 
            \\
            &=& \frac{2}{(\Delta p)_s \ r} {\Delta p}_s(r) = \frac{\mathcal{R}_s}{\gamma_s \ r}\ {\Delta p}_s(r),
        \end{eqnarray}
        and its $n$-th moment as
        \begin{eqnarray}\label{n-mom-s}
            \left\langle r^n \right\rangle_s &=& \frac{2}{(\Delta p)_s} \int_0^\infty r^{n-1} {\Delta p}_s(r) \ \mathrm{d}r 
            \nonumber \\
            &=& -\frac{1}{(\Delta p)_s} \int_0^\infty r^n \frac{\mathrm{d}p_N(r)}{\mathrm{d}r} \ ,
        \end{eqnarray}
        where $n \in \mathbb{Z}$.
        The normalization factor is, essentially, the pressure difference between the phase inside the spherical region (cluster / droplet / bubble) and the surrounding ambient phase, at distances far from the interface:
        \begin{eqnarray}
            (\Delta p)_s = -\int_0^\infty \mathrm{d}r \ \frac{\mathrm{d}p_N(r)}{\mathrm{d}r} \ .
        \end{eqnarray}
    
        The 3rd-moment of $f_s(r)$ trivially yields the cube of the radius of tension, $\langle r^3\rangle_s = (\mathcal{R}_s)^3$. 
        The radius of the interface of tension properly reads:
        \begin{eqnarray} \label{R_s}
            \mathcal{R}_s &=& 
            \left[-\frac{1}{(\Delta p)_s}\int_0^\infty r^3 \frac{\mathrm{d}p_N}{\mathrm{d}r} \ \mathrm{d}r\right]^{1/3} 
            \nonumber\\
            &=& \left[\frac{\displaystyle \int_0^\infty r^3 \frac{\mathrm{d}p_N}{\mathrm{d}r} \ \mathrm{d}r}{\displaystyle \int_0^\infty \frac{\mathrm{d}p_N}{\mathrm{d}r} \ \mathrm{d}r}\right]^{1/3}
            \nonumber\\
            &=& \left[\frac{\displaystyle \int_0^\infty \mathrm{d}r \ r^2 \ \Delta p_s(r)}{\displaystyle \int_0^\infty \mathrm{d}r \ r^{-1} \ \Delta p_s(r)}\right]^{1/3}
            \ ,
        \end{eqnarray}
        and, after imposing the {Young--Laplace (YL)} equation, yields the final expression of Thompson et al. for the surface (interfacial) tension~\cite{thompson1984molecular}:
        \begin{eqnarray}\label{delta_p_s}
            \gamma_s &=& \frac{(\Delta p)_s \  \mathcal{R}_s}{2} 
            \nonumber\\
            &=& \left[ -\frac{{(\Delta p)_s}^{2}}{8} \int_0^\infty r^3 \ \frac{\mathrm{d}p_N}{\mathrm{d}r} \ \mathrm{d}r \right]^{1/3} \ . \ \
            \ \ \ \ \ 
            \nonumber\\
            &=& -\frac{1}{2 (\mathcal{R}_s)^2} \int_0^\infty r^3 \ \frac{\mathrm{d}p_N}{\mathrm{d}r} \ \mathrm{d}r \ .
        \end{eqnarray}
        One can additionally notice that:
        \begin{itemize}
            \item The Bakker--Buff equation\cite{buff1955spherical,thompson1984molecular} and the integral condition for mechanical stability\cite{rowlinson2013molecular, thompson1984molecular} can be expressed as $\Gamma_{n=2}(r) = \gamma_s$ and $\Gamma_{n=-1}(r) = \gamma_s$, respectively.
            \item In both expressions, the radius of tension acts as the symmetry constant: $\Lambda_s = \mathcal{R}_s$ in Eq.~\ref{eq:Gamma_mom} for $n=-1$ and $n=2$.
            \item Also:
            \begin{equation}
            -\int_0^\infty r^3 \ \frac{\mathrm{d}p_N}{\mathrm{d}r} \ \mathrm{d}r = (\Delta p)_s \ (\mathcal{R}_s)^3 \ .
        \end{equation}
        \end{itemize}


\clearpage
\section{Extra results}\label{app:B}
\begin{figure}[h!]
    \centering
    \includegraphics[width=0.5\linewidth]{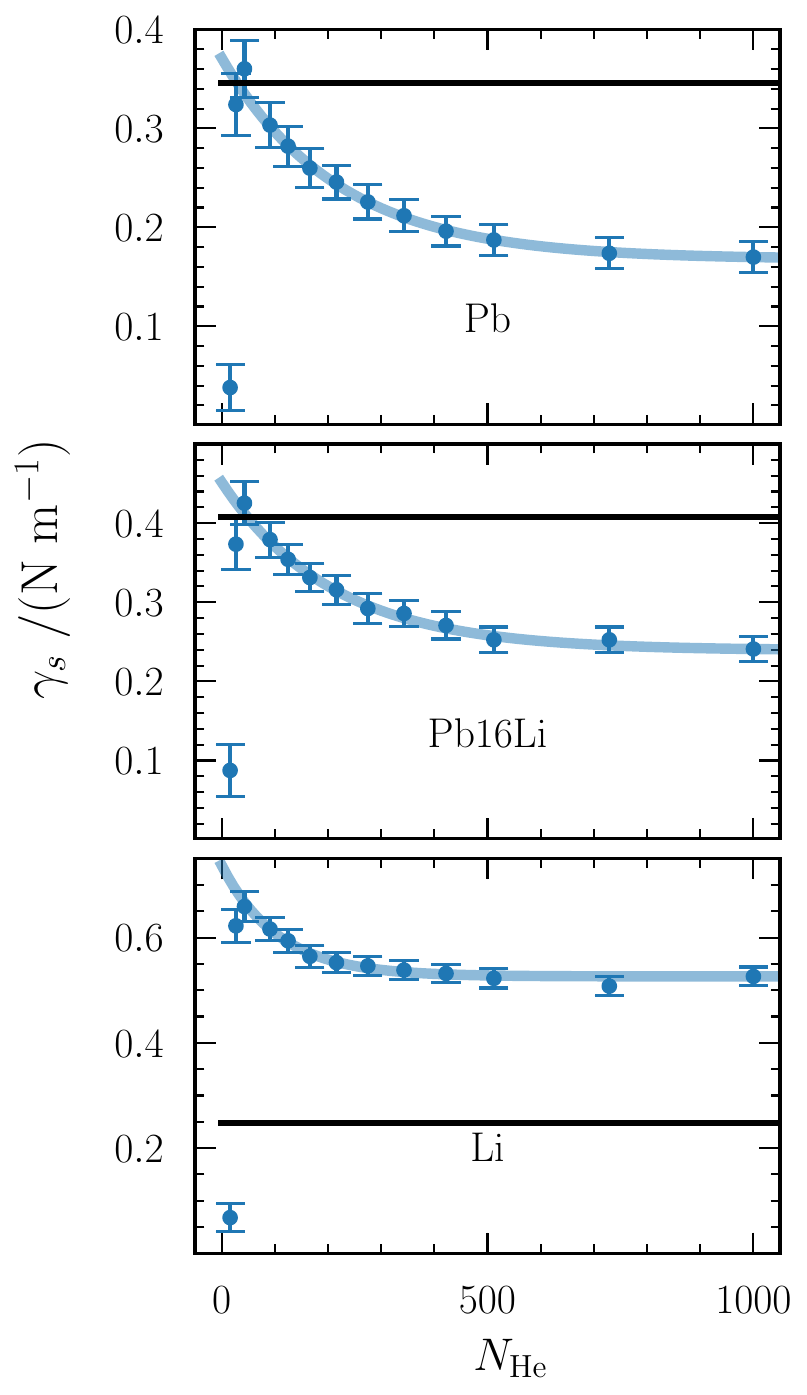}
    \caption{Interfacial tension values for varying helium cluster sizes at constant temperature (1021~K). Simulations have been performed using the Al-Awad force field (2023)~\cite{al2023parametrization}.}
    \label{fig:convergence_Awad_1021K}
\end{figure}

    In Figure~\ref{fig:convergence_Awad_1021K} the convergence of interfacial tension values at $T \approx 1021.4$~K for pure Li, pure Pb and Pb16Li solvents has been additionally depicted.

\begin{figure*}
    \centering
    \includegraphics[width=\textwidth]{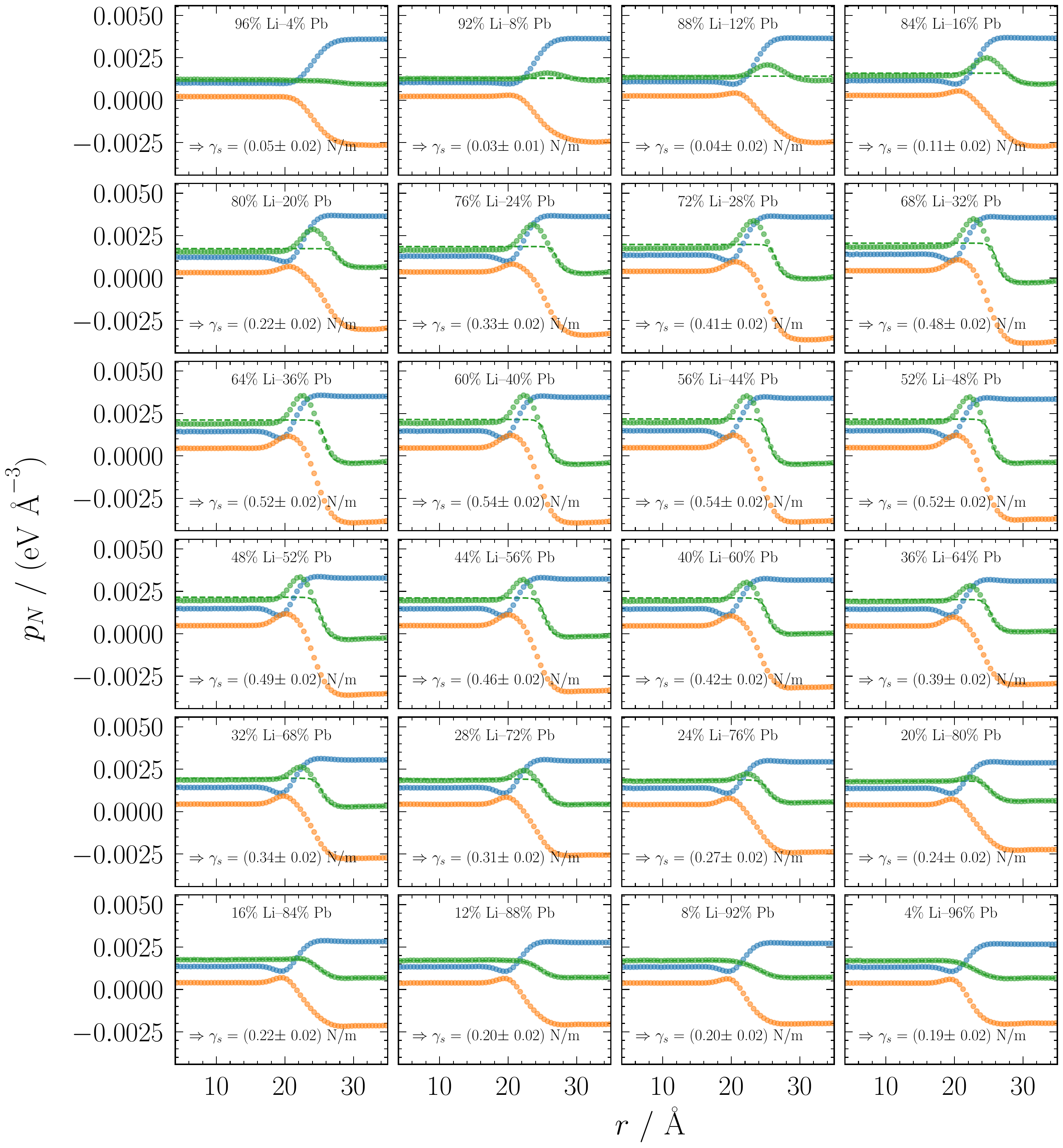}
    \caption{Normal component of pressure tensors in the radial direction for helium bubbles of $N_{\rm He}=512$ atoms in 24 different lead--lithium alloys (AEAM model~\cite{al2023parametrization}). Kinetic ($p_K$), virial ($p_V$) and total ($p_N=p_K+p_V$) profiles are depicted using blue, orange and green circles, respectively. In addition, $p_N^{id}$ have been estimated by fitting the corresponding $p_N$ values, and drawn using green dashed lines.
    }
    \label{fig:pNr_512_A}
\end{figure*}

\begin{figure*}
    \centering
    \includegraphics[width=\textwidth]{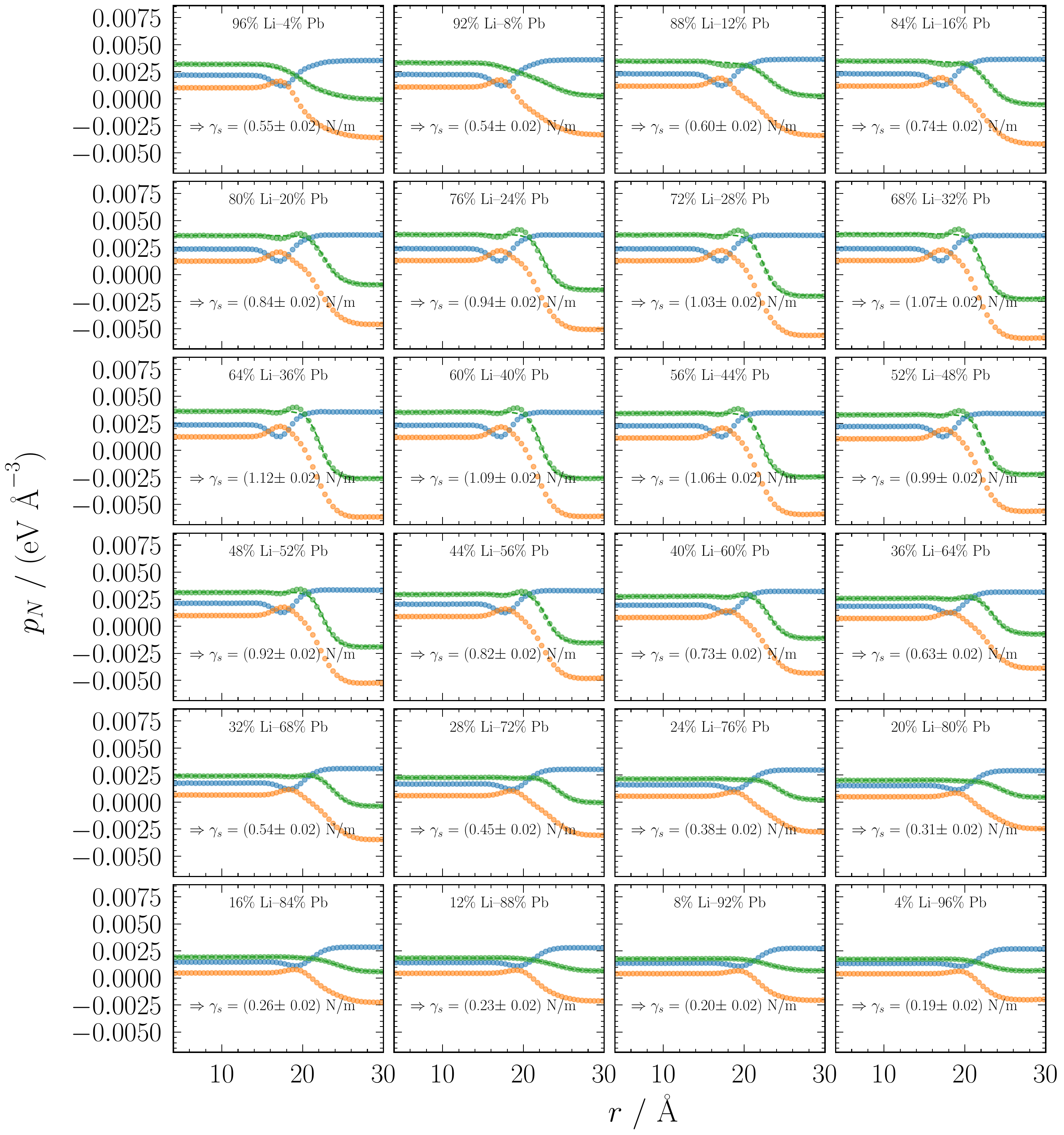}
    \caption{Normal component of pressure tensors in the radial direction for helium bubbles of $N_{\rm He}=512$ atoms in 24 different lead--lithium alloys (BEAM model~\cite{belashchenko2019inclusion}). 
    Kinetic ($p_K$), virial ($p_V$) and total ($p_N=p_K+p_V$) profiles are depicted using blue, orange and green circles, respectively. In addition, $p_N^{id}$ have been estimated by fitting the corresponding $p_N$ values, and drawn using green dashed lines.}
    \label{fig:pNr_512_B}
\end{figure*}

In Figures~\ref{fig:pNr_512_A} and \ref{fig:pNr_512_B} additional pressure tensors obtained at the same temperature are reported.


\end{document}